\documentclass[linenumbers]{aastex631}
\usepackage{epsfig,amsmath,amsfonts,amssymb,graphicx,subfigure,enumitem,color}
\usepackage{amssymb}
\usepackage{amsmath}
\usepackage[varg]{txfonts}
\usepackage{url}             % For breaking URLs easily through lines
\usepackage{graphicx}
\usepackage{float}
\newcommand{\beq}{\begin{equation}}
\newcommand{\eeq}{\end{equation}}
\newcommand{\beqa}{\begin{eqnarray}}
\newcommand{\eeqa}{\end{eqnarray}}
\newcommand{\beqann}{\begin{eqnarray*}}
\newcommand{\eeqann}{\end{eqnarray*}}
\bibpunct{(}{)}{;}{a}{}{,} % to follow the A&A style

\usepackage{textcomp}
\newcommand{\foo}{\rlap{+}{\texttimes}}
\shortauthors{Mishra et al.}
\shorttitle{Forced Magnetic reconnection}
\begin{document}
\nolinenumbers
\title{Formation of Jet-driven Forced Reconnection Region and Associated Plasma Blobs in a Prominence Segment}
\author[0000-0003-2129-5728]{Sudheer K.~Mishra}
\affil{Astronomical Observatory, Kyoto University, Kitashirakawa-Oiwake-cho, Sakyo-ku, Kyoto, 6068502, Japan}
\affil{Indian Institute of Astrophysics, Koramangala, Bangalore-560034, India.}
\author[0000-0002-1641-1539]{A.K.~Srivastava}
\affil{Department of Physics, Indian Institute of Technology (BHU), Varanasi-221005, India}
\author[0000-0003-0003-4561]{S.P.~Rajaguru}
\affil{Indian Institute of Astrophysics, Koramangala, Bangalore-560034, India.}
\author[0000-0002-7208-8342]{P.~Jel\'inek}
\affil{University of South Bohemia, Faculty of Science, Department of Physics, Brani\v sovsk\'a 1760, CZ -- 370 05 \v{C}esk\'e Bud\v{e}jovice, Czech Republic.}

\bigskip

\begin{abstract} 

{We use data from the \textit{Atmospheric Imaging Assembly (AIA)} onboard the \textit{Solar Dynamics Observatory (SDO)} to study the most likely formation of a forced reconnection region and associated plasma blobs, triggered by jet-like structures in a prominence segment. Around 05:44 UT on December 16$^{th}$, 2017, hot jet-like structures lifted from a nearby active region and fell obliquely on one side of the prominence segment with velocities of $\approx$45--65 km s$^{-1}$. These eruptions compressed the boundaries of the prominence and flux rope, forming an elongated reconnection region with inflow velocities of 47--52 km s$^{-1}$ and 36--49 km s$^{-1}$ in the projected plane. A thin, elongated reconnection region was formed, with multiple magnetic plasma blobs propagating bidirectionally at velocities of 91--178 km s$^{-1}$. These dense blobs, associated with ongoing reconnection, may also be linked to the onset of Kelvin-Helmholtz (K-H) instability. The blobs are attributed to plasmoids, moving at slower speeds (91--178 km s$^{-1}$) due to the high density in the prominence segment. The dimensionless reconnection rate varied from 0.57--0.28, 0.53--0.26, and 0.41--0.20, indicating reconnection rate enhancement and supporting the forced reconnection scenario. After reconnection, the prominence plasma heated to 6 MK, releasing significant thermal energy ($\approx$5.4$\times$10$^{27}$ erg), which drained cool prominence plasma and heated it to coronal temperatures. The ubiquity of jets and outflows in the solar atmosphere makes the aforementioned of reconnection and possible co-existence of K-H instability potentially important for the magnetic energy release and heating in the solar atmosphere.}
 
\end{abstract}

\section{Introduction}

Solar prominences are clouds of cool and dense plasma suspended against gravity within a magnetic field in a million-degree hot corona. They show a variety of dynamical plasma processes such as instabilities, waves, turbulence, bidirectional and counter-streaming flows, and oscillation \citep[e.g.,][and references cited therein]{2010SSRv..151..243L, 2010SSRv..151..333M, 2014LRSP...11....1P, 2018RvMPP...2....1H, 2018ApJ...856...86M, 2019ApJ...874...57M, 2021ApJ...923...72M}. Magnetic reconnection and instability may be two major causes of the eruption of solar prominnces. Magnetic reconnection is a crucial physical process in astrophysical and laboratory plasma. It is defined as the breaking and reconfiguration of two oppositely directed magnetic field lines. During this process, the stored magnetic energy converts into heat, kinetic energy, and radiation. Several theoretical studies have been conducted to understand the activities of small-to large-scale magnetic structures in the solar atmosphere, such as solar flares, jets, spicules, filament eruptions, coronal mass ejection, and space plasma \citep[e.g.,][]{1995ApJ...451L..83S, 1995Natur.375...42Y, 1999Ap&SS.264..129S, 2000mrmt.conf.....P, 2010RvMP...82..603Y, 2011A&A...535A..95D, 2011LRSP....8....1C, 2011LRSP....8....6S, 2020JGRA..12525935H, 2021SSRv..217...38K}. However, direct observation of magnetic reconnection in the solar atmosphere remains elusive. High-resolution data obtained from SDO, IRIS, and ground-based observatories (SST) provide a unique opportunity to observe and study the magnetic reconnection in different magnetic structures in the solar atmosphere \citep{2008ApJ...689L..69S, 2012SoPh..275....3P, 2014SoPh..289.2733D}. Most solar activities, such as small-to large-scale eruptions and different plasma processes in the solar atmosphere, such as jets, prominence, CMEs, solar flares, spicules, Ellerman bombs, and bright points, are governed by magnetic reconnection. Reconnection is also responsible for the heating of magnetized plasma, and it may drive the flows at different spatio-temporal scales in the solar atmosphere. The onset of magnetic reconnection may occur in two ways: \textit{viz.} spontaneous or forced. The theory and observations of spontaneous magnetic reconnections have been well studied. It is a major candidate for coronal heating \citep[e.g.,][]{1988ApJ...330..474P, 2010RvMP...82..603Y}, triggering eruptions \citep[e.g.,][]{2012NatCo...3..747Z, 2014Sci...346C.315P, 2015NatCo...6.7598S}, prominence eruptions \citep[e.g.,][]{2016NatPh..12..847L, 2016NatCo...711837X, 2020SoPh..295..167M}, jets \citep[e.g.,][]{1995Natur.375...42Y, 1997Natur.386..811I, 2015Natur.523..437S, 2015A&A...581A.131J, 2021NatAs...5...54A, 2023ApJ...945..113M}, the evolution of bright points \citep[e.g.,][]{2014Sci...346A.315T, 2014Sci...346C.315P}, spicules \citep{2019Sci...366..890S}, plasma flows \citep{2023ApJ...953...84M}, and solar flares \citep[e.g.,][]{1994Natur.371..495M, 2013NatPh...9..489S} under solar atmosphere.
 \\

In the lower solar atmosphere, the small-scale magnetic reconnections is in general initiated by the flux emergence \citep{1992PASJ...44..265S}, and release of photospheric shearing \citep{1992PASJ...44..265S, 2018ApJ...858L...4X}. Several small-scale features have also been identified using the high-resolution SST and IRIS data (e.g., UV bursts, Ellerman bombs, quiet Sun Ellerman bombs (QSEB), bright points, etc). Small-scale magnetic reconnection may be responsible for the evolution of these magnetic structures, which increases emitted radiation by several orders and governs the plasma flow of 100 km s$^{-1}$ at a smaller spatial scale (a few hundred km) in the lower solar atmosphere \citep{2014Sci...346C.315P}. Also, a small-scale magnetic reconnection was reported in the quiet region of the Sun and sunspot penumbrae and identified as an Ellerman Bomb \citep{2020A&A...641L...5J, 2021A&A...648A..54R}. It is also responsible for triggering ultraviolet bursts \citep{2014Sci...346C.315P}. These structures appear as bright needle-like structure in the high-resolution SST data, and magnetic reconnection is found to be the major cause of small-scale magnetic structures in the solar atmosphere. The small-scale magnetic reconnection may be associated with the nanoflares like energy range and can contribute to the coronal heating. It appears in various magnetic structures in the solar atmosphere, e.g., chromospheric jets \citep{2007Sci...318.1591S}, magnetic braids \citep{2013Natur.493..501C}, coronal loops \citep{2014Sci...346B.315T}, X-ray jets in coronal holes \citep{2015Natur.523..437S} using high-resolution observations covering from visible, UV, EUV, and X-ray data. High-resolution data of Extreme Ultraviolet Imager \citep{2020A&A...642A...8R} onboard the Solar Orbiter mission \citep{2020A&A...642A...1M} reveals the loop-like elongated brightening known as a EUV brightening, extreme-ultraviolet bursts and nanoflares in the quiet-Sun transition region and corona \citep{2021A&A...656L...4B, 2021A&A...647A.159C}. Magnetic reconnection is one of the acceptable mechanisms to trigger these EUV brightening in the solar atmosphere. Recently, bi-directional jets and nanoflares were also discovered in solar prominence, which are triggered due to the reconnection between ambient field and sheared prominence field \citep[e.g.,][]{2021A&A...651A..60H, 2021NatAs...5...54A}. As discussed in the above section, small to large-scale eruptions and magnetic structures are governed by spontaneous magnetic reconnection, which is well-studied theoretically and observationally. Moving further a leap forward, a new physical scenario is recently envisioned \citep{Sri24}, and numerically modelled \citep{2024ApJ...977..235M}, where a wave-like perturbation results in the collapse of a null and formation, reconnection and plasmoid dynamics in the current sheet further leading an onset of fast magnetoacoustic waves in the solar corona. This is termed as "Symbiosis of WAves and Reconnection (SWAR)". \\

The second method of magnetic reconnection involves the formation of a current sheet in an MHD stable configuration, where the reconnection process is perfectly seeded \citep[e.g.,][]{1985PhFl...28.2412H, 2017JPlPh..83e2001V, 2019A&A...623A..15P, 2024ApJ...963..139M}. The theory and modeling of forced reconnection are well developed both analytically and numerically \citep[e.g.,][]{1985PhFl...28.2412H, 1992PhFlB...4.1795W, 1998PhPl....5.1506V, 2001PhPl....8..132B, 2005GeoRL..32.6105B, 2005PhPl...12a2904J, 2017JPlPh..83e2001V, 2019A&A...623A..15P}. Energy release and plasma heating due to forced reconnection are derived analytically by \citet{1998PhPl....5.1506V}. Forced reconnection occurs even in MHD stable configurations when external deformation or perturbation forces oppositely directed magnetic fields to form a current sheet. The dynamical solar atmosphere may launch such perturbations in the form of sheared photospheric field, flux emergence, wave activities, or external eruptions, etc. As forced reconnection proceeds, the elongated current sheet breaks into multiple magnetic islands, similar to the tearing mode evolution \citep[e.g.,][]{2005GeoRL..32.6105B, 2005PhPl...12a2904J, 2015PhPl...22d2109C, 2015PhPl...22i0707V, 2017JPlPh..83e2001V, 2019A&A...623A..15P, 2024ApJ...963..139M}. A time lag appears between external perturbations and the internal response of the system before the onset of forced reconnection. This temporal lag is useful for determining the energy conversion and dissipation rate \citep[e.g.,][]{2005PhPl...12a2904J, 2017JPlPh..83e2001V, 2019A&A...623A..15P, 2019ApJ...887..137S, 2021ApJ...920...18S}. Although the analytical and numerical modeling approach of forced reconnection is well established, only a few observational studies have directly observed the onset of forced reconnection in the solar atmosphere \citep[e.g.,][]{2010ApJ...712L.111J, 2019ApJ...887..137S, 2019ApJ...879...74C, 2020A&A...643A.140M, 2021ApJ...920...18S}.  \\

It should be noted that there were significant differences between spontaneous and forced reconnection. Spontaneous reconnection may be associated with flux emergence \citep[e.g.,][]{1992PASJ...44..265S, 2014LRSP...11....3C} and the release of photospheric shear \citep[e.g.,][]{1988ApJ...330..474P, 2018ApJ...858L...4X}. The expansion of the eruptive field lines associated with flux emergence directly reconnects with the overlying coronal magnetic fields and further leads to magnetic reconnection. \citet{2016NatPh..12..847L} observed a driven-magnetic reconnection between an erupting filament and its nearby coronal loops, resulting in changes in the filament connection. The erupting filament-associated field lines were directly reconnected with the overlying coronal loops. The expanding coronal mass ejection may also be reconnected with the nearby active region \citep{2014ApJ...788...85V}. In addition, the misaligned internal magnetic field lines inside the prominence reconnect spontaneously, producing nanoflares in avalanche-like processes. There are no signs of external perturbations responsible for the misaligned internal magnetic field lines inside the prominence \citep{2021NatAs...5...54A}. Therefore, there is no time lag between the application of perturbations and the initiation of inflow/outflow in spontaneous reconnection, which is an important signature of forced magnetic reconnection, as suggested by theory and observations \citep[e.g.,][]{2005PhPl...12a2904J, 2017JPlPh..83e2001V, 2019A&A...623A..15P, 2019ApJ...887..137S, 2021ApJ...920...18S}. An additional step is involved in the forced reconnection process. External perturbations (e.g., eruptions, wave activities, flux emergence, displacement of photospheric footpoints, and coronal dynamics) may act as external drivers to perturb or force the surrounding or nearby magnetic field lines. The nearby field lines may reconnect with the overlying or nearby oppositely directed magnetic field. The external driver does not directly reconnect with the overlying or nearby fields in a forced magnetic reconnection. The difference between driven reconnection and forced reconnection was discussed in detail in a previous study on forced reconnection initiated by prominence \citep[e.g.,][]{2019ApJ...887..137S, 2021ApJ...920...18S}. \citet{2019A&A...623A..15P} applied recurring external perturbations with different strengths and showed that during forced reconnection, the current sheet broke into multiple magnetic islands through the tearing mode of instability. Recently, the Symbiosis of WAves and Reconnection is described in the solar plasma where the waves and reconnection processes show their livelihood in space and time, and their co-existence and physical properties highly depend on each other \citep{Sri24,2024ApJ...977..235M}. During this forced reconnection process, multiple magnetic islands are formed. They propagate bi-directionally along an elongated current sheet. Some magnetic islands merge with each other and form larger islands to enhance the reconnection rate. We would also like to highlight that such plasma blobs may also form via Kelvin-Helmholtz (KH) instability in different magnetized structures, such as solar jets \citep{2018NatSR...8.8136L, 2019SoPh..294...68S, 2021ApJ...923...72M}, prominences \citep{2010ApJ...716.1288B,2010SoPh..267...75R, 2017ApJ...850...60B, 2018ApJ...857..115Y, 2018ApJ...864L..10H, 2019ApJ...874...57M}, solar flares \citep{2018A&A...618A.135R, 2022ApJ...931L..32W} and CMEs \citep{2011ApJ...734L..11O, 2011ApJ...729L...8F, 2013ApJ...767..170F, 2013ApJ...766L..12M, 2015A&A...574A..55Z}, etc. Multiple studies and simulations have explored the simultaneous occurrence of KH instabilities and reconnections within various magnetized structures in the solar atmosphere. In regions where velocity and magnetic shear are present, a transformation occurs between the KH unstable vortices and resistive tearing mode (reconnected plasmoids) in a fan spine topology, as demonstrated by \citep{2013PhPl...20c2117W, 2021ApJ...923...72M}. Recent high-resolution observations also suggest the formation of plasma sheets via reconnection and multiple blob ejections via either reconnection or KH instability \citep{2021A&A...651A..60H}. Close coupling of the tearing mode and KH instability during turbulent magnetic reconnection has also been observed in the loop top of solar flares \citep{2023ApJ...954L..36W}. It was found that during the formation of the current sheet/elongated reconnection region, KH instability may evolve and may be responsible for the formation of larger structures (e.g., blobs or vortices) in the presence of shear flows. Such details of the physical processes have not yet been reported. \\
 \\
The present study elucidates about the formation of a thin elongated reconnection region and multiple propagating plasma blobs that may also likely associate with the onset of Kelvin-Helmholtz (K-H) instability. The highly dense and moving plasma blobs are detected in the thin elongated apparent current sheet formed during ongoing reconnection in the prominence segment. All these processes are observationally detected once the jet-like plasma structures fall on the prominence segment. The reconnection may start after the external coronal jet-like structures fall on the prominence segment. An elongated reconnection region-like field configuration formed at the outer periphery of the prominence segment may be subject to external disturbances to initiate reconnection. The prominence segment may already have internal twisted fields that may be pushed by external perturbations (as in the case of forced reconnection) to locally initiate reconnection in a current sheet-like structure. In addition, bidirectional flows in solar prominence were seen in EUV coronal lines using AIA data. The loop-like eruption contains multiple collimated jet-like structures. These multiple recurring jet-like structures act as external perturbations, which disturb the periphery of the prominence segment and initiate reconnection. The reconnection region and associated current sheet were fragmented into magnetic island chains, releasing their stored energy. Section~2 discusses SDO/AIA and STEREO/EUVI observations. Section~3 describes observational results of initiating magnetic reconnection by externally cooled jet-like structures. The discussion and conclusions are presented in section 4.\\

\begin{figure*}
  \includegraphics[trim = 0.0cm 0.0cm 0.0cm 0.0cm, scale=0.6]{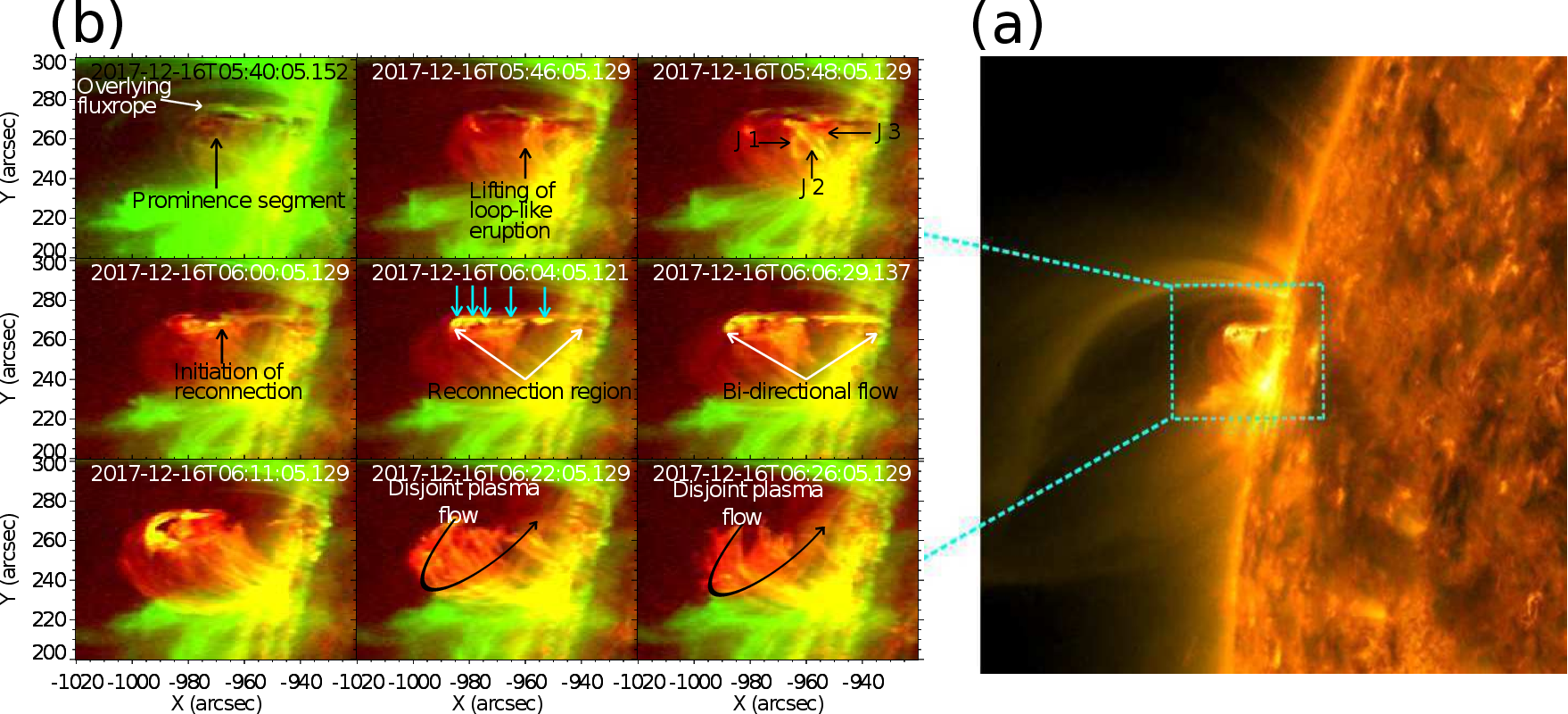}
\caption{The SDO/AIA 171+304 {\AA} composite image depicts a reconnection around 06:02 UT inside a prominence segment on 2017 December 16 (panel $'a'$). The region of interest (ROI) within the cyan dashed-line box is shown in zoomed images of panel $'b'$. This image sequence (panel $'b'$) displays a prominence segment, evolution of the reconnection region after the continuous bombardment of the multiple jet-like structures (J1, J2, J3) that may influence the boundary of the prominence segment and cause bi-directional plasma outflows inside the prominence. An overlying flux rope is also evident in this observed portion of the off-limb corona. The animation1.mp4 (171+304) {\AA} of the SDO/AIA composite images displays the onset of magnetic reconnection in the prominence  segment triggered externally by the jet-like eruptions and subsequent bi-directional flows evident as moving plasma blobs within the apparent current sheet therein. It runs from 05:30 to 06:40 UT. The black curvilinear path shows the disjoint plasma flow along the prominence body after the reconfiguration of the magnetic field takes place.}
\end{figure*}
%%%%%%%%%%%%%%%%%%%%%%%%%%%%%%%%%%%%%%%%%%%%%%%%%%%%
\vspace{1 cm}
\section{Observational Data and Analysis} 

The AIA \citep{2012SoPh..275...17L} onboard SDO \citep{2012SoPh..275....3P} consists of four telescopes, which capture 
the images of the Sun in ten different wavebands. Seven out of the ten filters are used for observing in the EUV wavebands: 94, 131, 171, 193, 211, 304, and 335 {\AA}, covering the regions from the upper chromosphere to the corona. The AIA/EUV channels have a spatial resolution of 1.5 arcsec, pixel width of 0.6 arcsec, and 12 seconds temporal cadence. On 2017 December 16, AIA observed some part of a solar prominence located near the 16$^{\circ}$ east limb (Figure~1). We use multi-channels of SDO/AIA data to understand the prominence dynamics, 
nearby jet-like eruptions, their impact on the prominence, and thereby formation and evolution of the reconnection region 
(Figures~1--4). The full prominence and its elongated channel appears in the STEREO-A/EUVI field of view 
\citep[FOV;][]{2004SPIE.5171..111W}. However, in SDO/AIA FOV, some part of the prominence is only visible, where 
reconnection is evident. \\

To understand the thermal structuring in and around the reconnection region, we perform differential emission measure 
(DEM) analyses using the method of \citet{2015ApJ...807..143C}. We have chosen the temperature between log $T$(K)=5.0\,--\,7.5 
with 25 temperature bins at log $T$(K)=0.1 intervals to deduce the DEM. The DEM map is plotted over a temperature range 
of 5.5$\le$ log T$\le$7.2 to understand the thermal behavior of the prominence, bi-directional flows in the form of 
magnetic islands due to the reconnection, and associated elongated reconnection region (Figure~5; left panel). We 
find that multi-thermal plasma evolves during the reconnection, and most of the cool and dense prominence 
plasma has heated up to the coronal temperature; therefore, we exclude optically thick AIA 304 {\AA} emission from the 
DEM analysis.\\

\begin{figure*}
  \includegraphics[trim = 0.0cm 0.0cm 0.0cm 0.0cm, scale=0.35]{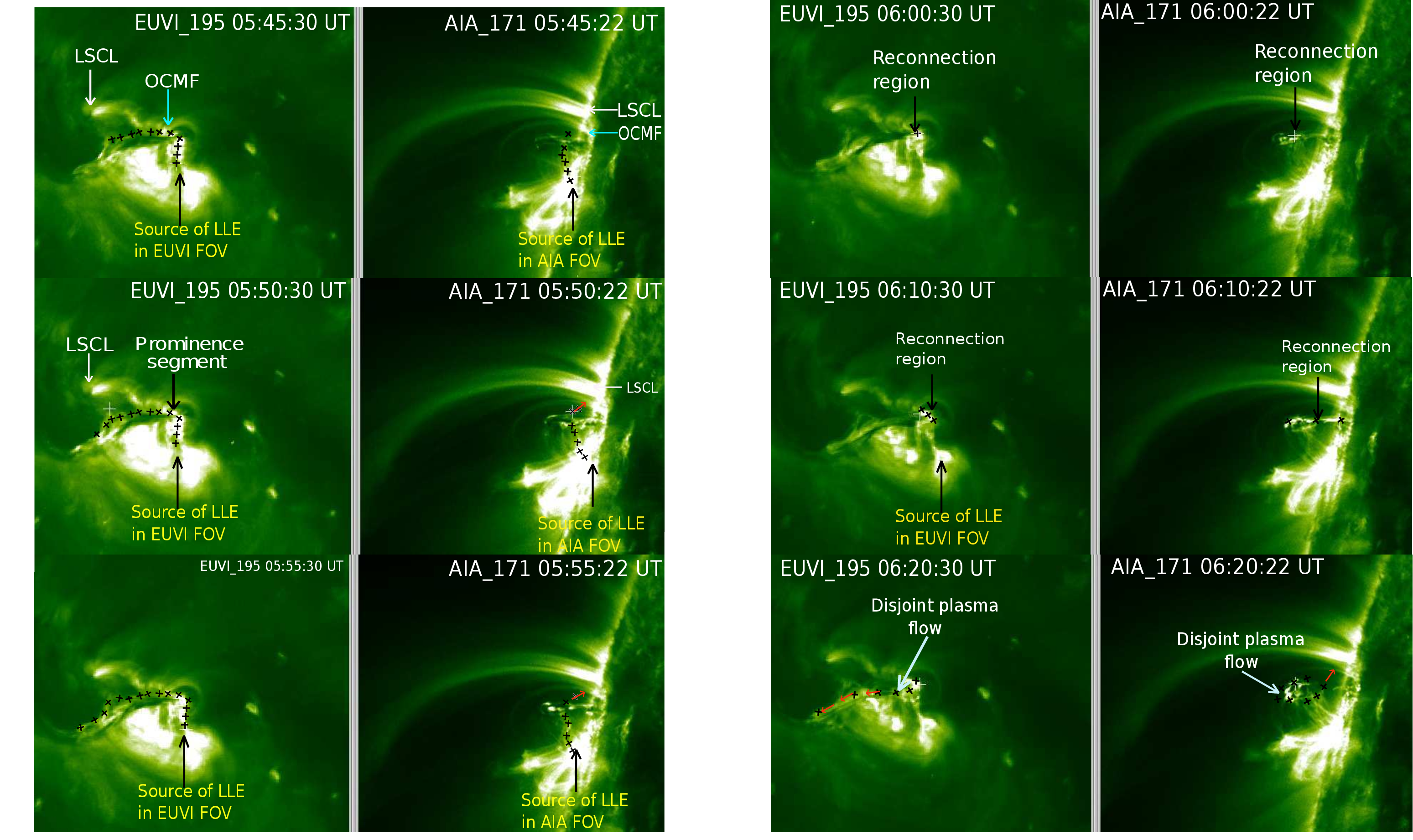}
\caption{This figure shows the source region of a loop-like eruption (LLE), the large-scale coronal loops (LSCL), an overlying coronal magnetic field (OCMF) configuration of the prominence segment, the reconnection region, and its associated dynamics in SDO/AIA 171 {\AA} and STEREO-A/EUVI 195 {\AA}. We use the tie-pointing method to locate the LLE, reconnection region, and other magnetized structures in the projected plane of EUVI image data. We tracked the different parts of the reconnection region (black + sign) in the AIA and EUVI field of view (FOVs) using the triangulation technique. We also track the possible source region of loop-like eruption (LLE; left panel) and disjoint plasma flows (lower right panel) after the completion of the reconnection process.}
\end{figure*}

\section{Observational Results}

On December 16, 2017, we observed a segment of quiescent prominence appearing near the northeast limb. It is surrounded by large-scale coronal loops and the associated magnetic field lines (Figure~1). The remaining part of the prominence lies beyond the visible part of the sun and does not appear in the AIA field of view. However, it is ovserved in the STEREO/EUVI field of view (FOV; Figure~2). This prominence lies near the active region AR 12692. Multiple collimated jet-like structures appeared inside the loop-like emerging structure and started to lift at $\sim$05:40 UT from the nearby active region AR 12692 (Figure~1, Animation1.mp4). Multiple collimated jet-like structures frequently originate from the active region starting at $\sim$05:44 UT inside the loop-like emerging eruption and fall on the prominence obliquely (Figure~1; top row of panel $'b'$, Animation1.mp4). However, in this study, we did not aim to understand the dynamic behavior of the active region, triggering of the jet-like structures, and their eruptions. Instead, we focus on the dynamics of the later phase when these jet-like structures obliquely push the prominence segment, mimicking the inflowing plasma structures pressing the outer periphery of the prominence-corona interface in its that segment. These jet-like structures (J1, J2, and J3; Figure~1; top row of  the panel $'b'$) transport bright plasma onto the prominence segment. The persistent motion of the plasma, manifested as coronal jet-like structures, exerts an external influence on the magnetic structures of prominences and potentially initiates magnetic reconnection. It is important to realize that the prominence segment and overlying magnetic flux rope may be situated in such a manner that the field lines are either in opposite directions or at some angles. The multiple jet-like collimated structures serve as an external perturbation that pushes inflowing plasma and causes the disruption of the region between the two magnetic structures, namely the prominence segment and the overlaying flux rope-associated field to initiate some reconnection. A reconnection region developed at the periphery of the prominence segment (Figure~1; middle row in panel $'b'$). The prominence segment breaks owing to the reconnection, bi-directional outflow starts, and multiple plasma blobs (akin of plasmoids/magnetic islands) propagate along the elongated reconnection region. Finally, after the reconnection, the prominence segment appears as a failed eruption (Figure~1, lower row in panel $'b'$; associated animation). A flux rope structure seems to have existed above this prominence. The eruption did not pass through the overlying corona because of the overlying field. It should be noted that the accumulated dense and hot plasma moves inside the prominence segment body, similar to the bright blob structures. However, the reconfiguration was initiated at $\approx$06:00 UT. After the initiation of the reconnection, we were able to observe the ejection of multiple bidirectional blobs (middle row in panel $'b'$; Figure~1, Animation1.mp4). As reconnection proceeds, multiple such plasma blobs most likely resembling the plasmoids are ejected, and subsequently, the apparent reconnection region (apparent current sheet) reconfigures. One segment of the reconfigured magnetic field is connected from the disk, while the most prominently associated plasma flows along the curvilinear path (disjoint plasma flow, black curvilinear path in the bottom row of panel $'b'$ of Figure~1, Animation1,mp4) in the main body of the prominence, which is observed in the STEREO-A/EUVI data. \\

%%%%%%%%%%%%%%%%%%%%%%%%%%%%%%%%%%%%%%%%%%%%%%%%%%%%%%%%%%%%%%%%%%%%%%%%%%%%%%%%%%%%%%%%%%%%%%%
\begin{figure*}
  \includegraphics[trim = 0.0cm 0.0cm 3.0cm 0.0cm, scale=0.7]{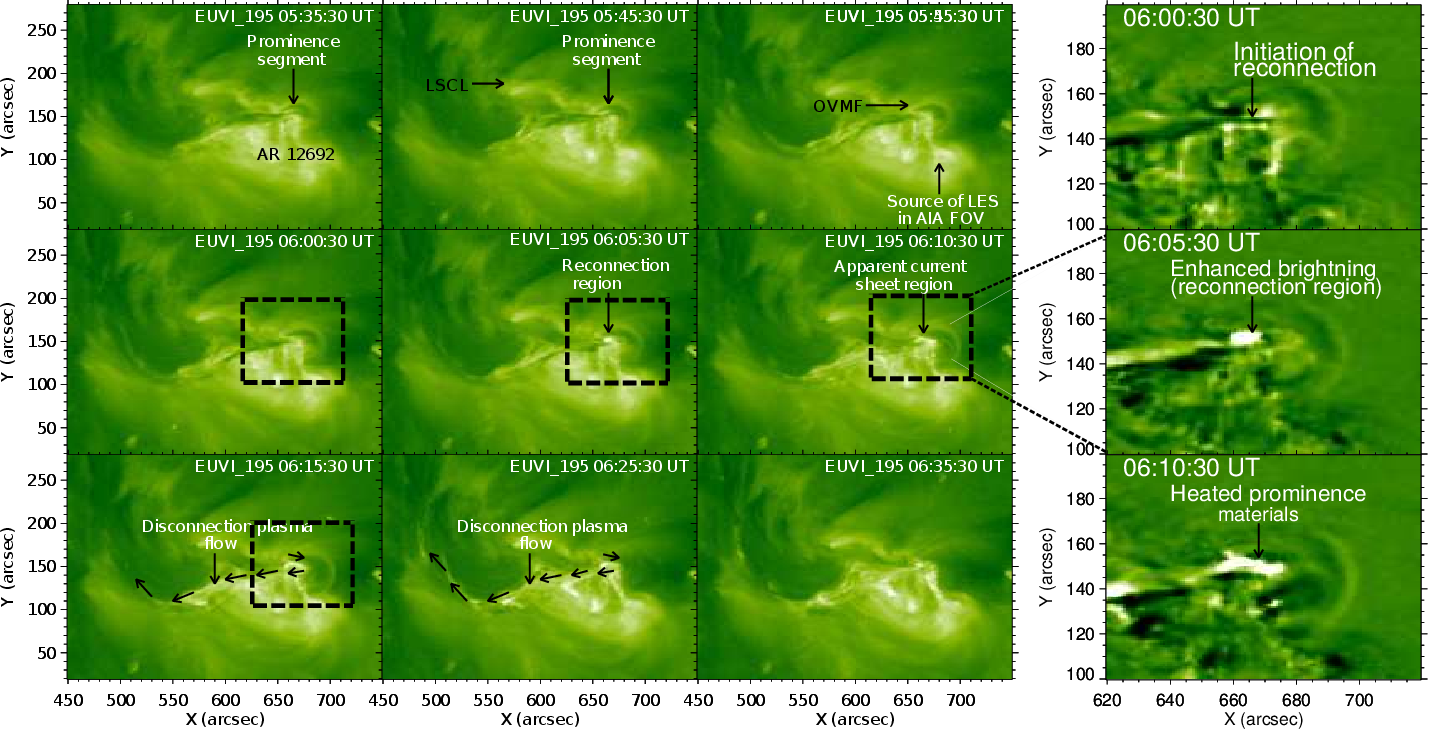}
\caption{ The loop-like eruption (LLE), magnetic reconnection, and disjoint plasma flow in the STEREO-A/EUVI 195 {\AA} data on December 16, 2017, are depicted in this picture, along with their temporal progression. In various panels, black color arrows have represented various magnetized structures, such as large-scale coronal loops (LSCL), overlying coronal magnetic fields, prominence segments, reconnection regions, active region AR 12692, source regions of loop-like eruptions (LLE), and disjoint plasma flow. Right panel: The likely forced reconnection region is visible on the disk in the zoomed view (black box region in the left panel) of the STEREO-A/EUVI 195 {\AA} running difference pictures. After the breakup of the elongated reconnection region ($\sim$06:10 UT), we observe an enhancement in brightness (during reconnection $\sim$06:05 UT) and the initiation of disjoint plasma flow as the reconnection begins.}
\end{figure*}
%%%%%%%%%%%%%%%%%%%%%%%%%%%%%%%%%%%%%%%%%%%%%%%%%%%%%%%%%%%%%%%%%%%%%%%%%%%%%%%%%%%%%%%%%%%%%%%%%%%%%%

\subsection{Tracking of Reconnection Region Using Tie-Pointing Method in Stereoscopic View}
  On December 16, 2017, the separation angle between SDO/AIA and STEREO-A was $\approx$123$^{\circ}$. It should be noted that in the active region, AR 12692, an extended prominence/filament system, appeared completely on the disk in the EUVI-A FOV. To obtain a better understanding of the large-scale coronal loops (LSCL), reconnection region, source region of loop-like eruptions (LLE), and overlying coronal magnetic field (OCMF), we used the tie-pointing method to compare the 3-dimensional views from AIA and STEREO-A/EUVI (Figure~2). These two spacecraft, SDO/AIA, and STEREO-A/EUVI, provide two different views of this failed  eruption and the associated externally governed magnetic reconnection. We selected one point in the SDO/AIA data (black + sign) and obtained the projected epipolar plane from the STEREO-A/EUVI data (Figure~2). Using the scc\_measure.pro routine (available in Solarsoft), we obtain the line-of-sight ray, which belongs to the SDO/AIA data, and its back-traced ray on the same epipolar plane provides its location in the STEREO-A/EUVI data (right panel, Figure~2). The intersection of these two lines lies on the same epipolar plane, and hence, it will provide a possible 3-dimensional view of the prominence segment, reconnection region, and associated dynamics.\\

\subsection{Stereoscopic View of the Reconnection Region}
 We used a series of STEREO-A/EUVI 195 {\AA} images to observe the spatio-temporal evolution of the prominence segment in the projected plane. The reconnection region, lifting of jet-like structures, elongated prominence/filament, associated channel, and pre and post-reconnection scenarios are indicated by black arrows in the STEREO-A/EUVI-195 FOV (Figure~3). We used a tie-pointing method \citep{2006astro.ph.12649I} to extract the possible location of the extended apparent current sheet in the STEREO-A field of view. The tracking method and its related details are mentioned in the previous subsection. Initially, we tracked larger magnetized structures such as large-scale coronal loops (LSCL) and overlying coronal magnetic fields (OCMF) in the AIA FOV and traced their location in the STEREO-A/EUVI FOV (Figure~2). The prominence segment appears as a vertical structure (therefore, we mostly see the top part of the prominence segment) in the projected plane of the STEREO-A/EUVI images (top panel of Figure~2). The source region of the loop-like eruption (LLE) is located near the boundary of AR 12692. This LLE contained multiple collimated bright structures (defined as jet-like structures). These collimated jet-like structures bombarded and fell on the one segment of prominence at different times (they best appear in the AIA FOV, and their back-traced location is shown in the STEREO-A/EUVI FOV; see Figure~3). Some parts of the LLE interact with the vertical prominence segment (see the accumulation of bright plasma in Figure~1, and associated animation1.mp4), while the rest of the plasma moves away in the main body of the prominence/filament that appears best in the EUVI 195 {\AA} FOV (see the black arrow in the left panel of Figure~3 in EUVI images). Using the tie-pointing method, we tracked the possible path of LLE and its interaction with the prominence segment in the projected plane (see Figure~2 left panel). The middle panel of Figure~3 shows the initiation of the reconnection driven by collimated jet-like structures in the projected plane. We identified a possible reconnection region in the projected plane (black box region; Figure~3). We tracked the top, middle, and bottom parts of the elongated apparent current sheet in the AIA FOV, and then traced its location in the STEREO-A/EUVI FOV using the tie-pointing triangulation technique (black + sign in the right panel of Figure~2). As reconnection initiates and progresses, enhanced bright structures (representing the possible reconnection region; right panel of Figure3) become visible in the running difference images of the EUVI 195 {\AA} waveband in the projected plane. Following the disconnection of the reconnection region, disjoint plasma flows are observed within the STEREO-A/EUVI and AIA fields of view (indicated by the black "+" sign with a red arrow in the bottom-right panel of Figure~2 and the black arrow in the left panel of Figure~3). In Subsection 3.3, we discuss the kinematical properties of the jet-like structures, including inflow and outflow dynamics during and after the initiation of reconnection inside the prominence segment.
\\

%%%%%%%%%%%%%%%%%%%%%%%%%%%%%%%%%%%%%%%%%%%%%%%%%%%%%%%%%%%%%%%%%%%%%%%%%%%%%%%%%%%%%%%%%%%%%%%%%%%%%%%%%%
\begin{figure*}
  \includegraphics[trim = 0.0cm 0.0cm 3.0cm 0.0cm, scale=0.4]{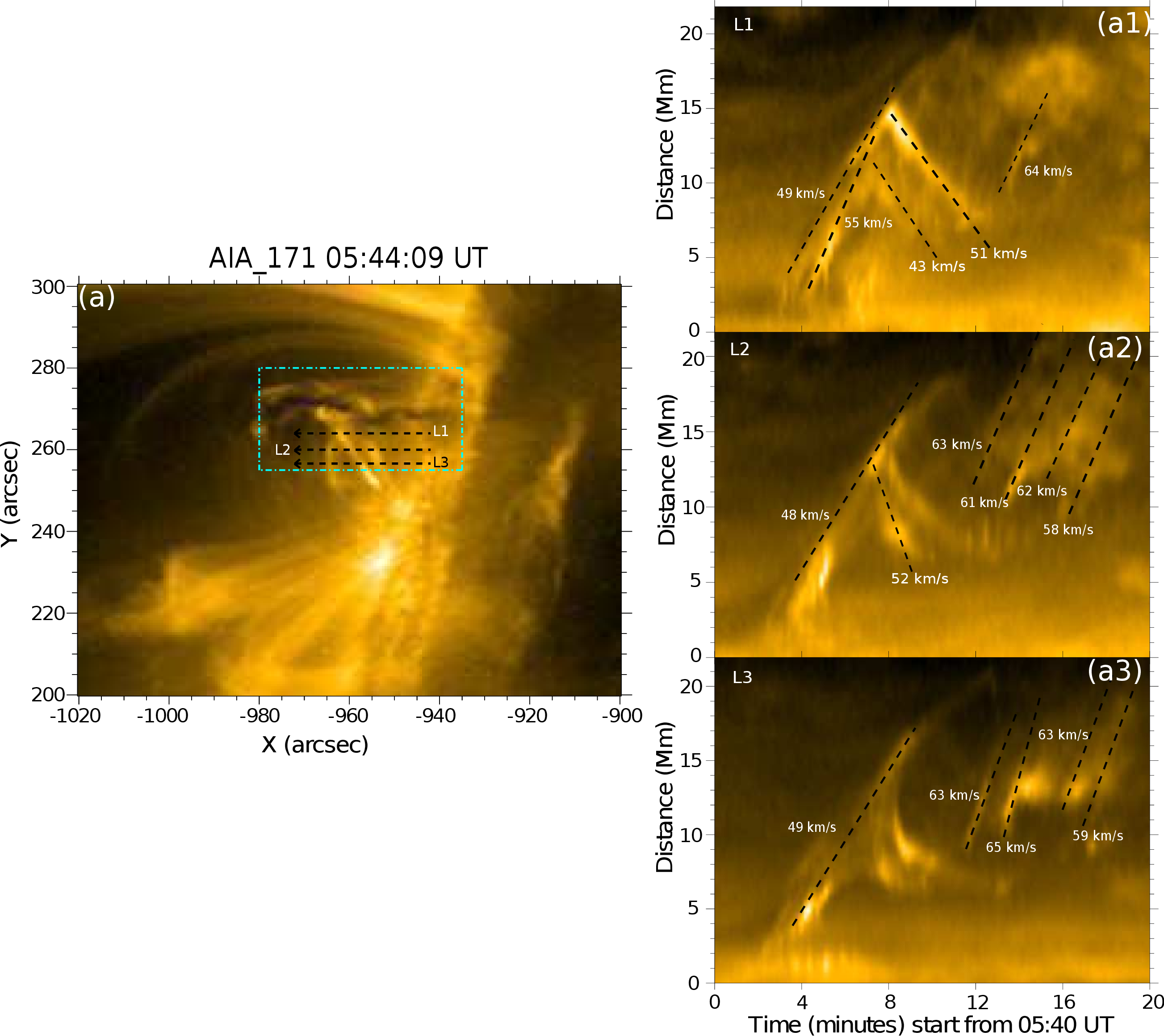}
\caption{ Left Panel: The intensity image of AIA 171 {\AA} is use to derive the height-time diagram, to undesrstand the kinematics of the lifting jet-like eruptions. Three slits 'L1', 'L2', and 'L3' are placed across the jet-like eruptions to track them at different heights (middle top panel). Multiple jet-like structures started to lift with a velocity of $\approx$49-65 km/s (right panel). 
They impinge on the nearby prominence obliquely and act as an external perturbation and might trigger the forced 
magnetic reconnection. Some plasma materials also fall along the length of prominence with a velocity of $\approx$43--52 km s$^{-1}$}
\end{figure*}

%%%%%%%%%%%%%%%%%%%%%%%%%%%%%%%%%%%%%%%%%%%%%%%%%%%%%%%%%%%%%%%%%%%%%%%%%%%%%%%%%%%%%%%%%%%%%%%
\subsection{Kinematical Properties}
The loop-like emerging structure began to rise from the nearby active region (top row in panel $'b'$ of Figure~1; animation1.mp4), containing multiple collimated bright structures. These collimated bright structures, resembling jet-like structures, collectively acted as external drivers, forcing the southward segment of the prominence-associated field lines to initiate reconnection within the prominence segment. We took three parallel slits (L1, L2, and L3) obliquely along the lifting jet-like structures to track them at different heights and times (panel $'a'$ of Figure~4). The right vertical column in Figure~4 displays time-distance plots and projected velocities towards the reconnection region of multiple jets covered by three slits, namely 'L1', 'L2', and 'L3' (panels $'a1-a3'$ of Figure~4). The slits used for generating the time-distance diagram have a width of 11 pixels, with 5 pixels on each side of the slit axis. The length of the slits extends to 50 pixels. The slits were positioned obliquely across the jet-like structures, and the estimated projected velocities ranged from $\approx$48--64 km s$^{-1}$ (panels $'a1-a3'$ of Figure~4). Initially, these jet-like structures lifted obliquely to fell on a part of the prominence, and were responsible for accumulating bright plasma within it. Some plasma downflows and the accumulation of bright plasma also occur along these collimated plasma flows, which are identified as jet-like structures. These downflows of the plasma also track on the same artificial slits (L1, L2, and L3). The estimated downflow velocity along the prominence is 41--52 km s$^{-1}$ (panels $'a1-a2'$ of Figure~4).\\

Figure~5 shows the onset of the reconnection region and associated plasma inflow dynamics. It occurs around $\approx$05:58 UT owing to the forcing of plasma inward by the jet-like structures associated with the field lines indicated by black arrows (top row of panel $'a'$ in Figure~5), bright accumulated plasma inside the prominence shown by the blue arrow (top and middle row of panel $'a'$ in Figure~5), and the overlying fluxrope indicated by red arrows (top and middle row of panel $'a'$ in Figure~5). It should be noted that the loop-like eruption-associated perturbations (which contain jet-like structures) act as an external driver to perturb the boundary (prominence-corona interface) of the twisted prominence segment and associated overlying flux rope fields in such a way that some favorable reconnection conditions appear locally. The collapse of these three magnetic field lines close to the reconnection area led to the initiation of the reconnection process (bottom row in panel $'a'$ of Figure~5). We placed an artificial slit $'S1'$ (having a width of 5 pixels and lenght of 25 pixels) to measure the inflow velocity in the reconnection region associated with the inward motion of the jet-driven magnetized plasma, which may push one boundary of the prominence segment and associated twisted field lines (indicated by the black arrow; middle row in panel $'a'$ of Figure~5) and overlying large-scale coronal loops related to the magnetic field (red arrow; middle row in panel $'a'$ of Figure~5). The inward movement of these field lines along path $'S1'$ is initiated at $\approx$05:58:40 UT with a velocity $\approx$47--52 km s$^{-1}$ (top row in panel $'b'$ of
 Figure~5) in the plane of the sky to trigger the reconnection. We observed that the three magnetized plasma chunks collapsed, initiating magnetic reconnection. Consequently, we placed an artificial slit, labeled $'S2'$ (blue slit; panel $'a'$ of Figure~5),  along the accumulated, jet-like structure-driven, bright, and dense plasma within the prominence (indicated by the blue arrow in the left panel of Figure5), as well as the overlying large-scale coronal field lines (red arrow in Figure~5). We extracted the projected inflow velocity of 36--49 km s$^{-1}$ starting at the same time $\approx$05:58:40 UT (bottom row in panel $'b'$ of Figure~5). Magnetic reconnection is a 3-dimensional process, however, we use here 2-dimensional imaging data to extract the inflow velocity. In the present study, the complex magnetic field structure and projection effect may significantly affect the estimation of the inflow velocity. One noticeable feature is the complex plasma motion during the inflow or initiation of reconnection. The exact location of the reconnection point is difficult to observe because of the projection effect and spatial resolution limit of SDO/AIA (1.5 arcsec). The estimated inflow velocity was the highest in the projected plane.\\

%%%%%%%%%%%%%%%%%%%%%%%%%%%%%%%%%%%%%%%%%%%%%%%%%%%%%%%%%%%%%%%%%%%%%%%%%%%%%%%%%%%%%
\begin{figure*}
  \includegraphics[trim = 2.0cm 0.0cm 3.0cm 0.0cm, scale=0.31]{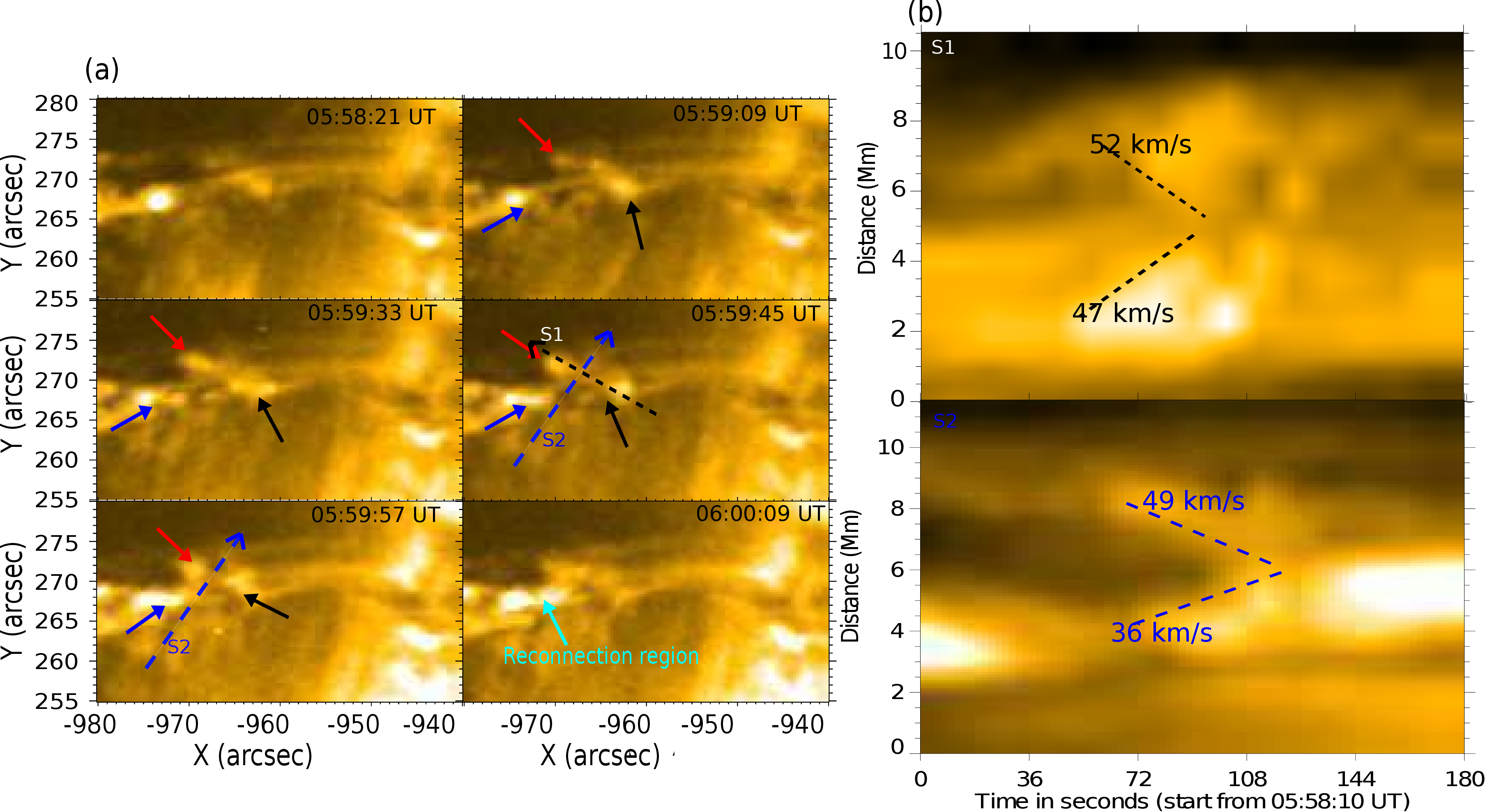}
\caption{ The intensity image of AIA 171 {\AA} uses the zoomed in view (cyan box in Figure~4) of the reonnection region, initiation of the reconnection. The reconnection associated inflows and field lines are indicated by blue, black, and red arrows (left panels). These field lines collapse with an inflow velocity of 47--52 km s$^{-1}$ along slit 'S1', while with a velocity of 36--49 km s$^{-1}$ along slit 'S2' at a point on the prominence segment and cause a reconnection (right panel; Figure~5). In various kinematical profiles, the length of the slits is indicated by the direction of the arrow within the slits.}
\end{figure*}
%%%%%%%%%%%%%%%%%%%%%%%%%%%%%%%%%%%%%%%%%%%%%%%%%%%%%%%%%%%%%%%%%%%%%%%%%%%%%%%%%%%%%%%%%%%%%%%%%%%%%%%%%%%%%%%

To investigate the kinematic properties and temporal evolution of the intensities of the reconnection region and estimate the bi-directional outflow velocity of the ejecting plasmoids following the onset of forced magnetic reconnection, we placed an artificial slit $'S'$ with a width of 11 pixels and a length of 120 pixels along the elongated reconnection region (panel $'a'$ of Figure~6). The temporal variations in the EUV intensity light curves, derived from the blue box shown in panel $'a'$ of Figure~6, are plotted for seven different wavebands (94, 131, 171, 193, 211, 304, and 335 {\AA}). This analysis aims to gain insight into the multi-thermal characteristics of the observed reconnection region and associated plamoids ejection (panel $'b'$ of Figure~8). It is important to acknowledge that we plotted the normalized intensity ($I/I_{max}$) to analyze and compare the multi-thermal characteristics of the plasma within the reconnection area. The commencement time of the reconnection process is denoted by a vertical black line in panel $'b'$. It appears that before the onset of reconnection, the normalized intensity in the light curve derived from various wavebands of the Atmospheric Imaging Assembly (AIA) remains relatively constant, except for a single instance at $\approx$05:44 UT (14$^{th}$ minutes; panel $'b'$ of Figure~8. The deposition of multithermal plasma via loop-like eruptions and collimated plasma structures (jet-like structures) is responsible for the increase in the intensity of the light curve. Following the aggregation of heated plasma, it initiates motion along the prominence segment (see animation1.mp4), subsequently leading to a slight reduction in the intensity of the light curve until 05:58 UT. As the reconnection proceeds, the rapid enhancement of temperature/heating in the reconnection region can be indirectly understood by measuring the intensity of different EUV filters of the AIA (panel $'b'$ of Figure~8). We found that the emissions in the cool and coronal AIA filters 304 {\AA} (log$T_{e}$= 4.7), 171 {\AA} (log T$_{e}$= 5.9), 131 {\AA} (log T$_{e}$=5.6, 7), 193 {\AA} (log $T_{e}$=6.1), and 211 {\AA} (log $T_{e}$=6.3) peaked between 06:05 UT--06:08 UT (in between 35--38$^{th}$ minutes in panel $'b'$ of Figure~8). Simultaneously, the intensities of the hot AIA filters 335 (log $T_{e}$=6.4) and 94 (log $T_{e}$=6.8) increased and peaked at approximately 06:12 UT (42$^{nd}$ minutes in the panel $'b'$ of Figure~8). This demonstrates a rapid energy release in the reconnection region over $\sim$12 min timescale (panel $'b'$, Figure~8). The prominence segment disappears as the cool and dense plasma material was heated to the coronal temperature constituted by the hot plasma (animation1.mp4). \\

It should be noted that the dynamical processes begin in the form of jet-like structures that fell on the prominence body with a velocity of 44--58 km s$^{-1}$ (panels $'a1-a3'$ of Figure~5). A time lag of $\approx$5 min appears between the external jet-like drivers (05:44--05:54 UT) and the initiation of the inflows ($\approx$05:58-06:00 UT). The reconnection was initiated at $\approx$06:00 UT, as indicated by the black and green vertical dotted lines (panel $'c'$ of Figure~6). Panel $'d'$ of Figure~8 shows a zoomed-in view of the time-distance diagram from 05:59--06:13 UT to track the outflowing bidirectional magnetic islands (green dotted vertical line in panel $'c'$ of Figure~8). After the onset of reconnection, multiple magnetic islands form, which propagate bi-directionally with a range of velocities between 91--178 km s$^{-1}$ as estimated along slit 'S' (panel $'d'$ of Figure~6). The velocity of the outflowing islands depends on the density of the reconnection region and the local magnetic field. Therefore, the estimated outflow velocity is smaller owing to the projection effect and the denser prominence plasma present in the reconnection region. The fundamental concept of reconnection theory suggests that the outflow speed of the ejecting plasma materials/jets is the Alfv\'{e}n velocity if the anti-parallel field lines reconnect with each other. The Alfv\'{e}n velocity was $\approx$1000 km/s at the typical coronal conditions \citep{2004psci.book.....A}. This shows that the estimated outflow speed in the present observations (93–-178 km/s) was the transonic speed at the coronal temperature \citep{2004psci.book.....A}. However, it should be noted that in the present study, reconnection was initiated in cool and dense prominence plasma materials. Therefore, the outflow velocity decreases. Some magnetic islands merge and form a larger island while propagating along the elongated reconnection region, as seen in the recent simulation by \citet{2024ApJ...977..235M}. The merging of magnetic islands may be governed by the coalescence instability \citep{2019A&A...623A..15P,2024ApJ...977..235M}. As the island coalescence proceeds, a thin elongated reconnection region is formed, leading to faster reconnection and rapid heating (Figure~6). \citet{2019A&A...623A..15P} applied multiple pulses with different strengths and found that multiple magnetic islands formed that propagated along the elongated reconnection region. Subsequently, these islands merged, increasing the reconnection rate. These magnetic islands evolve and propagate bi-directionally along an elongated reconnection region, likewise in a similar way as seen in a recent numerical simulation by \citet{2024ApJ...977..235M}. Some magnetic islands merge and form a larger island while propagating along the elongated reconnection region. In this study, we estimated the reconnection rate by considering the inflow and outflow ratios (R=$\frac{V_{inflow}}{V_{outflow}}$). Initially, the magnetic islands propagated at a lower velocity(91--103 km s$^{-1}$). However, as the reconnection proceeded, the estimated outflow velocity increased and decreased (panel $'d'$ in Figure~6; 91--129 km s$^{-1}$, 98--137 km s$^{-1}$, and 126--178 km s $^{-1}$). The decrease in velocity may be related to the merging of islands. The merged islands are larger and can propagate at a lower velocity. Using the estimated values of inflow (36--52 km s$^{-1}$) and outflow velocities in the projected plane, the estimated reconnection rate changes from 0.57-0.28, 0.53-0.26, and 0.41-0.20. \\
%%%%%%%%%%%%%%%%%%%%%%%%%%%%%%%%%%%%%%%%%%%%%%%%%%%%%%%%%%%%%%%%%%%%%%%%%%%%%%%%%%%%%%%%%%%%%%%
\begin{figure*}
  \includegraphics[trim = 0.0cm 0.0cm 0.0cm 0.0cm, scale=0.28]{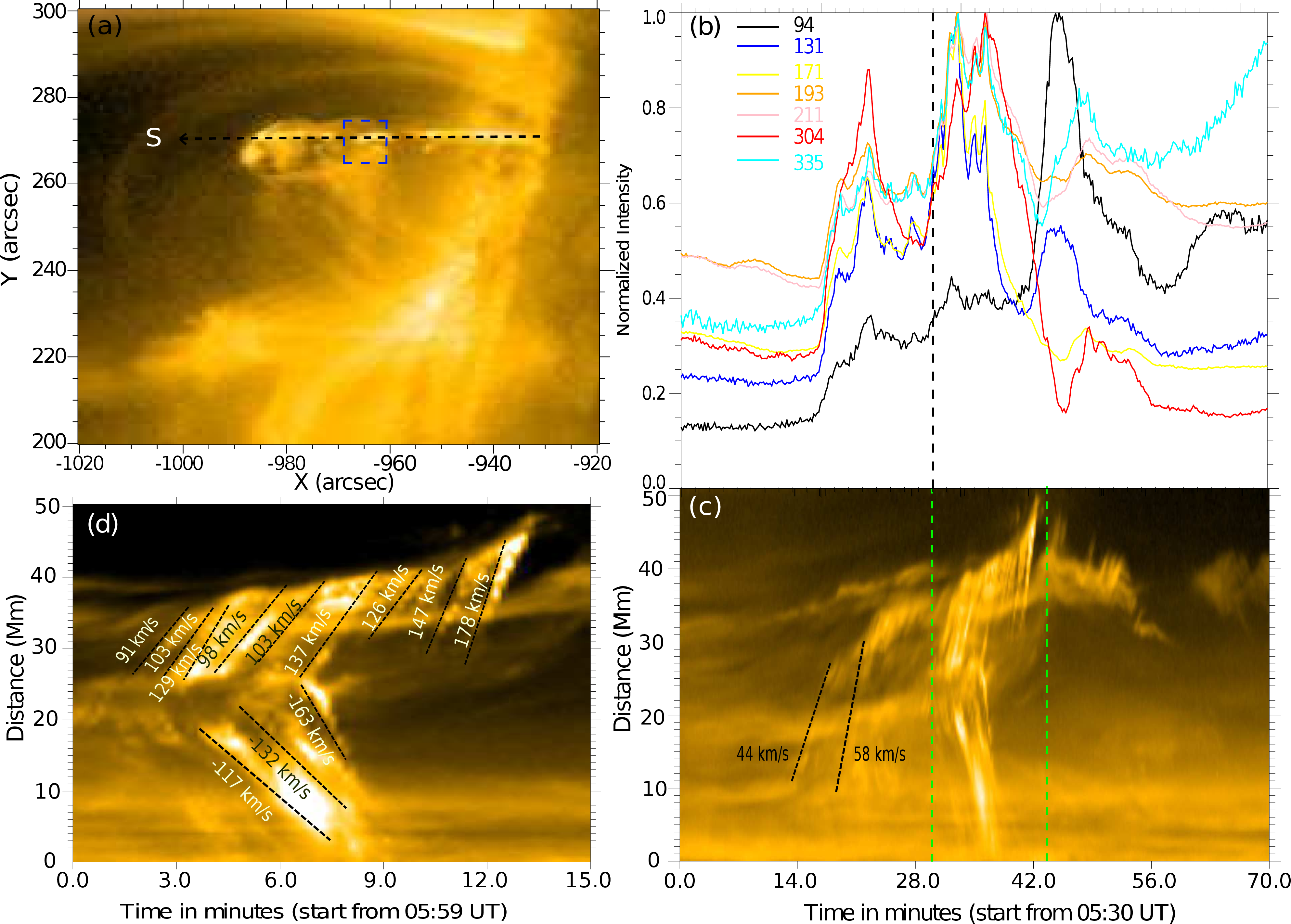}
\caption{The bi-directional flows of the magnetic islands observed along slit 'S' are shown in the bottom panels. The 
emissions recorded from the elongated reconnection region (blue-dotted box) in all EUV channels of AIA are shown in 
the top right panel.}
\end{figure*}
%%%%%%%%%%%%%%%%%%%%%%%%%%%%%%%%%%%%%%%%%%%%%%%%%%%%%%%%%%%%%%%%%%%%%%%%%%%%%%%%%%%%%%%%%%%%%%%%%%%%%%
 
Therefore, the decrease in the outflow velocity after the initiation of magnetic reconnection indicates that the reconnection rate increases. However, estimating inflow and outflow velocities in a projected plane is complicated. Small changes in the calculation of inflow and outflow velocities may significantly change the estimated reconnection rate. In addition, as the reconnection proceeded, multiple magnetic islands were formed in the elongated reconnection region (Figure~7). These islands formed in the middle of the apparent current sheet and propagated bidirectionally with an Alfv\'{e}nic velocity inside the cool and dense prominence plasma, which is essentially a low speed when we compare with the typical Alfv\'en speed in the solar corona. In the present study, the reconnection occurs within the cold, dense prominence plasma, with a typical plasma density of $\approx$10$^{11}$ cm$^{-3}$ and a temperature range of 8000--50000 K \citep{2010SSRv..151..333M}. Based on the findings of the aforementioned study, including the fact that reconnection is initiated by external perturbations, there is a time-delay between the application of perturbations and the generation of the plasma outflows. Also, the multiple magnetic islands are formed, and the coalescence instability increases the rate of the reconnection. All this collectively suggest that the aforementioned reconnection scenario proceeds as per the physical concepts of the forced reconnection.\\

%%%%%%%%%%%%%%%%%%%%%%%%%%%%%%%%%%%%%%%%%%%%%%%%%%%%%%%%%%%%%%%%%%%%%%%%%%%%%%%%%%%%%%%%%%%%%%%
\begin{figure*}
 \mbox{
  \includegraphics[trim = 0.0cm 1.0cm 0.0cm 0.0cm, scale=0.55]{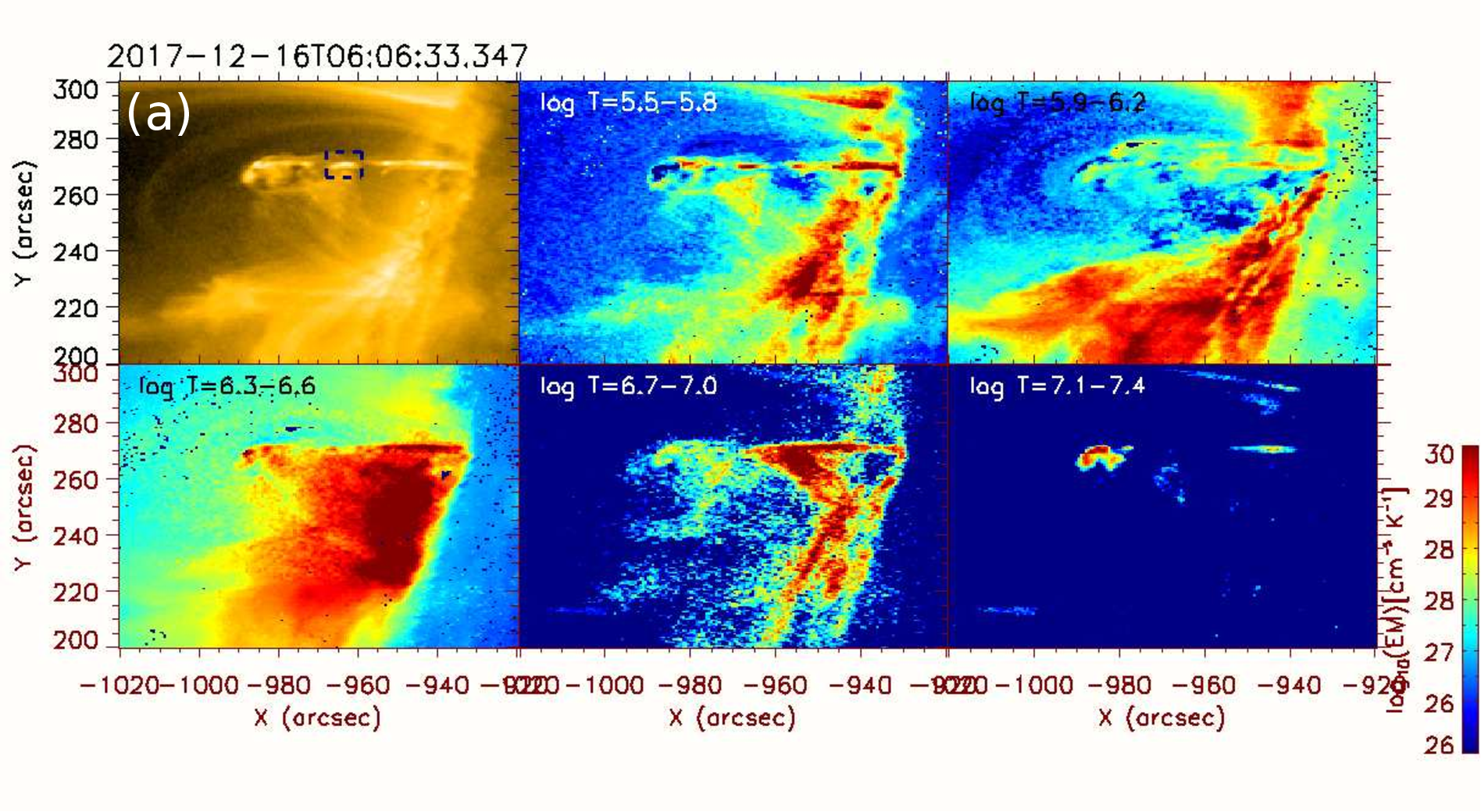}
  }
  \mbox{
  \includegraphics[trim = 1.5cm 0.0cm 0.0cm 0.0cm, scale=0.55]{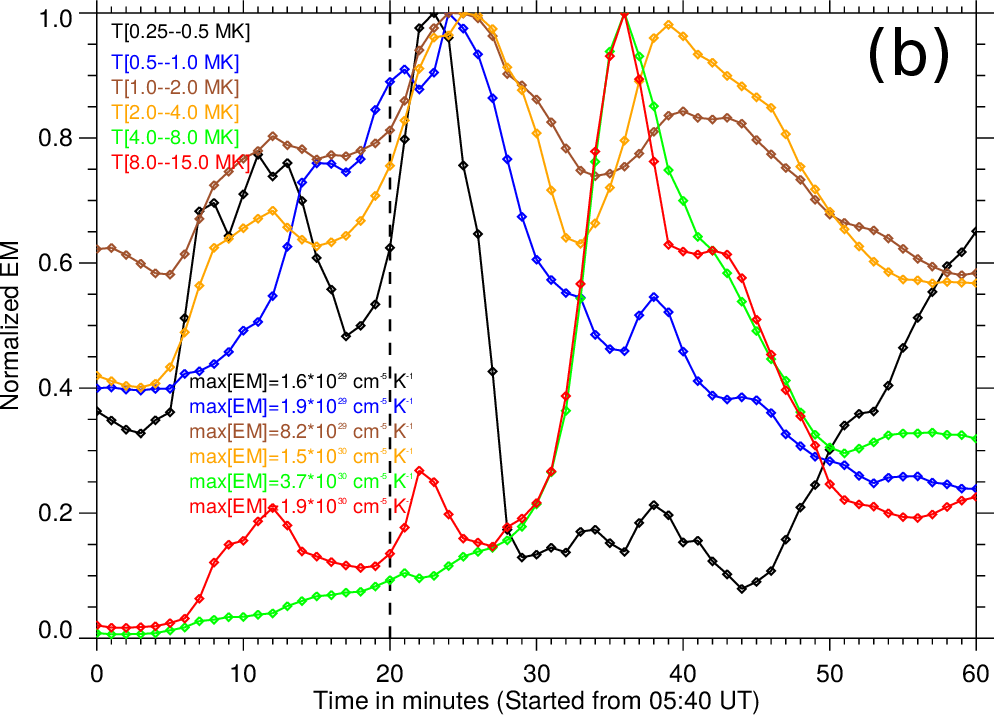}
    \includegraphics[trim = 0.0cm 0.0cm 0.0cm 0.0cm, scale=0.55]{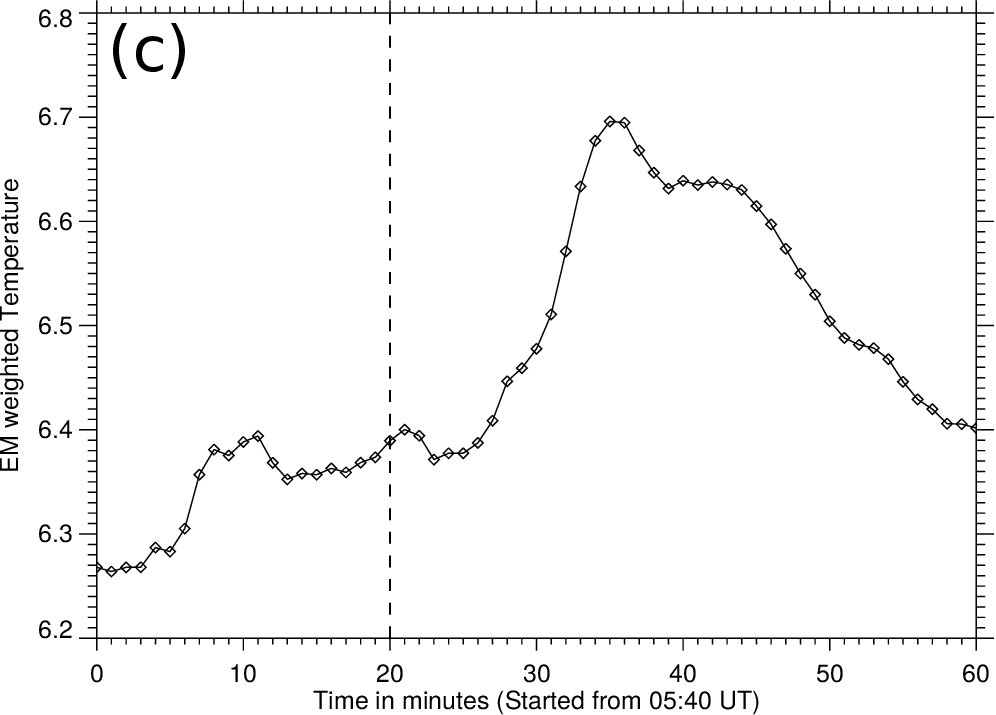}

  }

\caption{Panel (a) elucidates the DEM maps of reconnection region at a different temperature at log T$_{e}$=5.7--7.2. The number density of the elongated reconnection region is extracted from the blue box as shown in 171 {\AA}. Panel (b) displays the total emission extracted from the same blue box in 171 {\AA} at different temperatures at 
T=0.25-15 MK. Panel (c) shows the estimated EM-weighted temperature extracted from the blue box region. The vertical black dotted line indicates the onset time of the force reconnection $\sim$06:00 UT. The estimated average temperature and the emission measure extracted for the high-temperature bins [4-15 MK] peak simultaneously.} 

\end{figure*}
%%%%%%%%%%%%%%%%%%%%%%%%%%%%%%%%%%%%%%%%%%%%%%%%%%%%%%%%%%%%%%%%%%%%%%%%%%%%%%%%%%%%%%%%%%%%%%%%%%%%%%
Interestingly, the typical coronal Alfv\'{e}n speed in the active regions was $>$1000 km/s, whereas our calculated speeds were 91--178 km/s from the outflowing plasma inside the prominence. It is worth noting that after the disappearance of the cool prominence plasma, a significant amount of heating occurred, as observed during the peak intensity enhancement in the AIA 94 {\AA} channel (formed at 8 MK) at $\approx$06:12 UT (panel $'b'$ in Figure~6). This suggests that a portion of the stored magnetic energy is converted into the kinetic energy of the outflowing plasma, while the remaining magnetic energy is initially absorbed by the cool and dense prominence plasma and subsequently led to the significant heating. Previous research has also found similar results when magnetic reconnection occurs in cool, partially ionized, and collision-dominated plasma (i.e., in the photosphere, chromosphere, and prominence). The released magnetic energy was consumed by the surrounding cool and dense plasma \citep{1999ApJ...515..435L, 2001ChJAA...1..176C, 2006ApJ...641.1217C, 2019ApJ...887..137S, 2021ApJ...920...18S}. It is important to note that the outflow velocity of the plasma strongly depends on the surrounding environment and magnetic field structure. The outflow velocity can be either transonic or subsonic depending on the reconnection field within the prominence system \citep{2021A&A...651A..60H}. Interestingly, previous reports regarding the reconnection inside cool and dense prominence/filament plasma suggest that the outflow speeds can be transonic, subsonic, or Alfv\'{e}nic, for example, 40--170 km/s \citep{2016NatPh..12..847L}, 90--460 km/s \citep{2016NatCo...711837X}, 70--90 km/s \citep{2020SoPh..295..167M}, and 5--35 km/s \citep{2019ApJ...887..137S, 2021ApJ...920...18S}, depending on the reconnection field alignment in the prominence and surrounding cool material. \citet{2004ApJ...602L..61C} reexamined the inflow from the reconnection region and found that the apparent motion in the EUV images was not the actual inflow/outflow plasma motion. A detailed two-dimensional magnetohydrodynamic (MHD) simulation was conducted by \citet{1996PASJ...48..353Y} to evaluate various physical parameters associated with X-ray jets initiated through magnetic reconnection. Their findings indicated that the velocity of the magnetic islands/plasmoids is 0.1 times the outflowing Alfv\'{e}n velocity. In the present study, the actual outflowing plasma motion might be significantly high (Alfv\'{e}n velocity); however, here, we observe the apparent outflow plasma motion of the propagating magnetic islands/plasmoids inside the elongated reconnection region lying inside the cool and dense prominence segment. Therefore, the actual outflow may propagate with Alfv\'{e}n velocity (10$\times$extracted plasmoid velocity,i.e., $\sim$1000 km.s$^{-1}$). Therefore, to extract the actual plasma dynamics inside the reconnection region, a detailed spectroscopic analysis is required to estimate the 3D velocity field, internal dynamics of the apparent current sheet, and associated heating by these dynamical phenomena \citep{2005ApJ...622.1251L, 2006ApJ...648..712H, 2018ApJ...854..122W, 2022NatCo..13..640Y, 2023ApJ...945..113M}. Furthermore, magnetic islands/plasmoids resulting from reconnection also exhibit an enhancement in plasma density. The thermal properties, temperature, and density diagnostics resulting from reconnection inside the prominence segment are discussed in the upcoming Subsection~3.4.\\

%%%%%%%%%%%%%%%%%%%%%%%%%%%%%%%%%%%%%%%%%%%%%%%%%%%%%%%%%%%%%%%%%%%%%%%%%%%%%%%%%%%%%%%%%%%%%%%%%%%%%% 
\subsection{Thermal Properties}
To understand the thermal behavior and heating scenarios of the jet-driven perturbations, prominence segment, reconnection region, and associated dynamics, we performed a DEM analysis using six hot EUV filters of AIA (94, 131, 171, 193, 211, and 335 {\AA}). Initially, the overlying loops covered a cool and dense prominence plasma (Figures~1, 9, 10; animations~1,2,3). Jet-like plasma structures lift at $\approx$05:44 UT and impinge on prominence segment. They accumulated hot plasma inside the prominence segment from 05:44--05:54 UT (Figures~1,7; animation2.mp4). It should be noted that the accumulated bright plasma inside the prominence segment, due to the jet-like structures, consists of dense and hot plasma (Figures~7, 8; Animations 2, 3). Such external perturbations disturb the boundary of pre-existing prominence-associated twisted fields and overlying magnetic fluxrope-associated field lines in such a way that it starts a forced reconnection at $\approx$06:00 UT in the prominence segment. As reconnection progresses, the multi-thermal magnetic islands are formed inside the elongated reconnection region (Figure~7; animation2.mp4). We also plotted the temporal evolution of the total emission measure (EM=$\int_{T_{min}}^{T_{max}}{DEM(T) dT}$) and the average EM-weighted temperature ($T$=$\frac{\Sigma_{i=1}^{n}(DEM_{i} \times~logT_{i}}{\Sigma_{i=1}^{n}DEM_{i}}$) extracted from the reconnection sheet region (blue box region in panel $'a'$ of Figure~7). We distribute the entire temperature bins in the range of 0.25--15 MK  (panel $'b'$; Figure~7). We find that the total emission measure starts to increase for the low-temperature bins (0.25--4 MK) just before the initiation of reconnection (vertical black dotted line; panel $'b'$ of Figure~7). It peaked after the initiation of reconnection, that is,$\approx$06:04 UT. No signature of reconnection occurred before the onset of forced reconnection, as indicated by the vertical black dotted line (panel $'b'$; Figure~7). Interestingly, the temporal evolution of the total emission (EM) for the hot temperature bins (T=4--15 MK) did not significantly change (panel $'c'$ of Figure~7) up to 06:08 UT and then suddenly increased (please see the green and red curves in panel $'b'$ of Figure~7). The enhancement in the total emission for the hot temperature bins confirms some heat release during the reconnection and formation of the magnetic islands.\\
%%%%%%%%%%%%%%%%%%%%%%%%%%%%%%%%%%%%%%%%%%%%%%%%%%%%%%%%%%%%%%%%%%%%%%%%

However, it starts to increase significantly for the hot temperature bins T=4--8 MK and T=8--15 MK at $\sim$06:08 UT (28$^{th}$ minutes in panel 'b' of Figure~7). The total emission measure from the hot temperature bins (T=4--15 MK; green and red line curves in panel $'b'$ of Figure~7) increased more than 45--115 times after forced reconnection, indicating a rapid energy release and subsequent heating with a time lag of 8--12 min. Undoubtedly, the reconnection initiated at 06:00 UT within the cool and dense prominence plasma, resulting in the formation of an elongated reconnection region between 06:00 and 06:07 UT. Subsequently, the disjointedness of the reconnection region and the disappearance of the prominence plasma occurred. After this event, we observed the presence of heated plasma at the same location, with an average temperature of $\approx$6 MK (panel 'c' of Figure~7). It is important to note that most of the energy released after the reconnection was likely consumed by the cool and dense plasma. This energy release followed the merging of magnetic islands in the top and bottom parts of the prominence, which was also depicted by \citet{2019A&A...623A..15P}. To further confirm that the enhancement in temperature started after the initiation of the reconnection, we plotted the temporal variation of the EM-weighted temperature extracted from the same blue box region in panel 'a' of Figure~9. We found that initially (t = 05:40 UT), the extracted EM-weighted temperature was log T=6.28. The extracted temperature slightly increases after 05:45 UT and reaches up to the log T=6.40 (panel 'c' of Figure~7). This small enhancement in the EM-weighted temperature may be associated with the oblique hits of the jet-like structures and the accumulation of hot plasma in the prominence segment during the same time interval. It should be noted that the EM-weighted temperature started to suddenly increase at approximately 06:04 UT (just after the reconnection; see the vertical black line at 06:00 UT in panel 'c' of Figure~7) and peaked at approximately 06:15 UT. During the peak time of the EM-weighted temperature, its value reaches log T= 6.7, which clearly shows a significant amount of heat release after the initiation of forced magnetic reconnection with a time delay of $\approx$12--15 min (panel 'c'; Figure~7).\\

%%%%%%%%%%%%%%%%%%%%%%%%%%%%%%%%%%%%%%%%%%%%%%%%%%%%%%%%%%%%%%%%%%%%%%%%%%%%%%%%%%%%%%%%%%%

To investigate the heating processes within the prominence segment and the disappearance of prominence plasma resulting from jet-driven forced reconnection, we derived the temperature map from the emission measure (Figure~8). The temperature map was computed using the equation:
\begin{equation}
T(i) = \frac{\Sigma_{i=1}^{n}(DEM_{i} \times~T_{i})}{\Sigma_{i=1}^{n}DEM_{i}}
\end{equation}
where DEM(i) represents the differential emission measure and T(i) denotes the corresponding temperature. \\
In panels $'(a1)'$, $'(a2)'$, $'(a3)'$, and $'(a4)'$ of Figure~8, we present the temperature map and the temperature maps masked above 3 MK, 4 MK, and 5 MK, respectively, at t=06:06:09 UT. It is important to emphasize that three masking conditions were applied to exclude cooler plasma with temperatures below 3 MK, 4 MK, and 5 MK, effectively isolating and highlighting regions of hot plasma (Figure~8 and animation3.mp4). The temporal evolution of the dynamics is illustrated in animation3.mp4. This animation captures the eruption of the loop-like structure containing multiple collimated bright jet-like features with hot plasma, the accumulation of hot plasma within the prominence segment between 05:44 UT and 05:54 UT, the triggering of reconnection, the formation of fragmented hot plasma blobs (plasmoids) within the elongated forced reconnection region, the development of post-reconnection loops, and the heating of prominence plasma. We would like to highlight that the lifting loop-like eruption, containing multiple collimated jet-like structures between 05:44 UT and 05:54 UT, contains significantly hot plasma, which is clearly observable in the masked temperature map for temperatures exceeding 5 MK (animation3.mp4). These jets push the prominence-corona boundary, accumulating hot plasma with temperatures above 5 MK (refer to animation3.mp4). The average temperature profile extracted from the emission measure (panel $'c'$, Figure~7) shows a noticeable enhancement after 06:06 UT, reaching a peak of $\approx$6 MK at 06:12 UT. This observation aligns well with the corresponding temperature map. Additionally, We observe a significant increase in plasma associated with regions identified in the masked temperature map for temperatures exceeding 5 MK (Figure~8, animation3.mp4). Recent high-resolution 2.5-dimensional magnetohydrodynamic (MHD) simulations have investigated the energetics of coronal current sheets under the influence of asymmetric external perturbations. These studies have shown that, in the presence of resistivity, thermal conduction, and radiative cooling, the fragmentation of the current sheet leads to a reduction in the average magnetic energy density, while the average kinetic energy density increases \citep{2025ApJ...979..207M}. Our observations align with these findings, as we observed an increase in the velocities of the ejecting plasmoids (from 91 km s$^{-1}$ to 178 km s$^{-1}$; panel $'d'$ of Figure~6) during the progression of fragmentation. Additionally, we noted an enhancement in heating (thermal energy) as the fragmentation proceeds, with a peak occurring after the disjointment of the reconnection region (Figures~7, 8; animation3.mp4).\\

We measured the number density (n$_{p}$=$\sqrt{\frac{EM}{d}}$) within a magnetic island propagating along the elongated reconnection region (black box in panel $'a'$; Figure~8) by measuring the total emission. Substituting EM=5.5$\times$10$^{31}$ cm$^{-5}$ and  $'d'$=3.9 Mm (where $'d'$ is the depth that is equal to the width of the selected region), the estimated number density is n$_{p}$=3.8$\times$10$^{11}$ cm$^{-3}$, and the corresponding mass density $\rho_{p}$=8.2$\times$10$^{-13}$ g cm$^{-3}$. Taking the outflow velocity as Alfv\'en velocity (Figure~6), the strength of the reconnecting field line is $V_{A}\sqrt{4\pi\rho_{p}}$$ \sim$ 29--56 G. We adopted this method to estimate the thermal energy associated with a current sheet \citep{2021ApJ...908..213L}. The thermal energy generated in the reconnection region was TE=$\frac{3}{2}$. n$_{p}$.k$_{B}$.V.$\delta$T. Assuming that the elongated reconnection region is a cylinder, we estimate TE=$\frac{3}{2}$. n$_{p}$.k$_{B}$.$\pi(\frac{l}{2})^{2}$, respectively.L.($T_{2}-T_{1}$). Here $n_{p}$ is the number density of magnetic islands, $V$ volume, $k_{B}$ Boltzman's constant, and $\delta T$ is temperature increase ($T_{1}$) of prominence plasma to hot temperature plasma ($T_{2}$) of the reconnection region and magnetic islands. We estimated the temporal variation of the average weighted temperature from the reconnection region (blue box in panel $'a'$; Figure~7) to observe the evolution of the temperature during and after the reconnection. Before the onset of reconnection, the prominence plasma lying at a cool temperature was visible in the AIA 304 {\AA} filter. The estimated average weighted temperature is extracted from the box region before the initiation of reconnection, that is, $\approx$T$_{1}$=1.9$\times$10$^{6}$K (at 05:40 panel $'c'$ of Figure~7). After the onset of reconnection, the elongated reconnection region was heated to $\approx$T$_{2}$=6.0$\times$10$^{6}$K (at 06:15 UT; panel $'c'$ of Figure~7). Putting n$_{p}$=3.8$\times$10$^{11}$ cm$^{-3}$, $l$=0.7--1 Mm, L$\approx$28 Mm, T$_{2}$=6.0$\times$10$^{6}$K, and T$_{1}$=1.9$\times$10$^{6}$K, the calculated thermal energy is $\approx$5.4$\times$10$^{27}$ erg. The thermal energy is sufficient to heat the localized corona that contains the reconnection region. During the eruption of the emerging loop from the nearby active region, which hosts the jet-like structures that interact with the overlying loop arcade, cool prominence materials are hosted. During this event, localized brightening appeared in the loop arcades and prominence-associated plasma. The observed brightening phenomena could potentially be linked to the cumulative motion of the plasma and the interior dynamics of the prominences. An alternative hypothesis has been proposed, indicating that these brightening may be associated with nanoflares resembling nano jet activity, potentially releasing energy on the order of approximately 10$^{24}$--10$^{25}$ ergs \citep{2014Sci...346B.315T, 2021NatAs...5...54A}. Various magnetic structures of the sun have been documented to exhibit such phenomena. Nevertheless, it is important to acknowledge that even minor reconnection events on a tiny scale have not resulted in a substantial release of the energy required for abrupt heating in the solar prominence. Nevertheless, it is crucial to recognize that a recent extensive investigation of identical observations was conducted by \citet{2023ApJ...943..156K}, using data from SDO, STEREO, and IRIS. The results of their study suggest an alternative perspective. Researchers have determined that the observed nanojets are plasmoids that are in motion within an unsuccessful filament eruption.\\

%%%%%%%%%%%%%%%%%%%%%%%%%%%%%%%%%%%%%%%%%%%%%%%%%%%%%%%%%%%%%%%%%%%%%%%%%%%%%%%%%%%%%%%%%%%%%%%
\begin{figure*}
\begin{center}
  \includegraphics[trim = 0.0cm 0.0cm 0.0cm 0.0cm, scale=0.55]{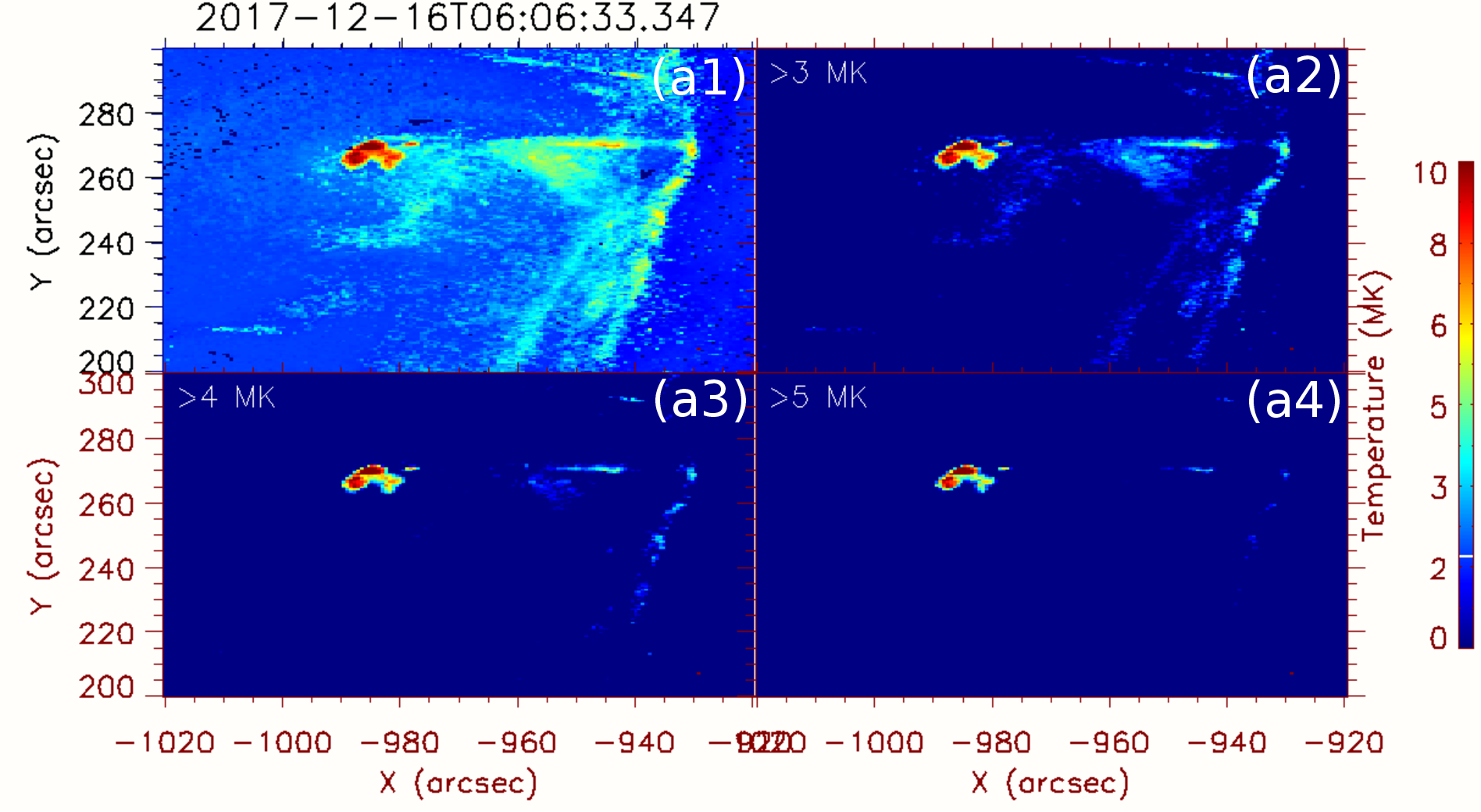}
\caption{Panels (a1), (a2), (a3), and (a4) display the complete temperature map, and temperature maps masked above 3 MK, 4 MK, and 5 MK, respectively, as extracted from the emission measure. The reconnection region, plasmoids, and associated dynamics appear in the hot temperature bins, confirming the formation of significant hot plasma following the onset of forced magnetic reconnection. An animation showing the temporal evolution of the emission measure-derived temperature map is available. This animation captures the overall dynamics, including the launch of loop-like eruptions containing multiple collimated jet-like structures, the accumulation of hot plasma, the initiation of reconnection, the formation of hot plasmoids, and the emergence of significant hot plasma material within the prominence segment. The events are shown between 05:40 UT and 06:30 UT, with a real-time duration of 12 seconds for the animation.}
\end{center}
\end{figure*}
%%%%%%%%%%%%%%%%%%%%%%%%%%%%%%%%%%%%%%%%%%%%%%%%%%%%%%%%%%%%%%%%%%%%%%%%%%%%%%%%%%%%%%%%%%%%%%%%%%%%%%
%%%%%%%%%%%%%%%%%%%%%%%%%%%%%%%%%%%%%%%%%%%%%%%%%%%%%%%%%%%%%%%%%%%%%%%%%%%%%%%%%%%%%%%%%%%%%%%%%%%%%%
We use AIA 131 {\AA} waveband to observe the appearance of the reconnection region, which consists of the ejection of multiple magnetic islands, a thin elongated current sheet, and hot post-reconnection loops (Figure~9; panel $'a'$). After the initiation of magnetic reconnection at $\approx$06:00 UT, we show the region of interest at t = 06:03:09 UT, after the proper development of the reconnection region, and multiple magnetic islands began to form within the thin, elongated plasma sheet/apparent current sheet (Figure~9; panel $'b'$). The zoomed-in and re-binned view of the reconnection region is obtained from the white box region in panel $'b'$ to observe the magnetic islands and the thin elongated apparent current sheet (panel $'c'$ in Figure~9). We found that bidirectional hot plasma blobs (akin to moving plasmoids) were ejected after the onset of the reconnection. A fragile and diffuse apparent current sheet/plasma sheet also appears in the zoomed FOV of the reconnection region (panel $'c'$ of Figure~9). Five slits were placed over five different locations (orange, pink, and magenta for plasmoids and dotted cyan and green lines for apparent current sheet) vertical to the reconnection region to estimate the width of the magnetic islands and the associated apparent current sheet (panel $'c'$ of Figure~9). We measured the variation in intensity along the length of all the five slits and displayed in panel $'e'$ of Figure~9. A Gaussian distribution (five different color curves) was used to estimate the width of the propagating magnetic islands and elongated current sheet. The average variation in intensity across these slits in the reconnection region estimates the average width of the magnetic islands by fitting a Gaussian distribution (solid black curve for magnetic islands and solid green curve for the apparent current sheet; panel $'e'$ of Figure~9). The full width at half maximum (FWHM) of this distribution is $\approx$7 pixels ($\sim$3 Mm) for the plasmoids/plasma blobs and $\approx$3 pixels ($\sim$1.3 Mm) for the thin apparennt current sheet (from green curve), which provides the average width of the magnetic islands and apparent current sheet respectively (panel $'e'$ of Figure~9). We adopted a similar process to estimate the width of the plasmoids/plasma blobs at t=06:04:09 UT by placing four additional slits (blue, yellow, red, and brown) and extracting the intensity variation along these slits (panels $'c', 'd'$; Figure~9). The vertical black lines on the two solid curves indicate the errors in extracting the intensity variations along these slits, calculated as the standard deviation of the distribution (panel $'e'$, Figure~9). We would like to emphasize that the current sheet cannot be resolved using the available AIA imaging channels. In the present paper, we also did not make any claims regarding the resolution of the current sheet and the associated internal dynamics, so we refer to it as an elongated reconnection region/apparent current sheet. It should be noted that the real current sheet may lie inside the elongated bright plasma structure/reconnection region. The propagation of bidirectional plasma magnetic islands, also known as plasmoids or plasma blobs, can be observed in all AIA channels. In conclusion, the AIA observable is the moving plasma blobs, however, physically they should be associated with the moving plasmoids/magnetic island formed due to the thinning and fragmentation of the reconnecting current sheet \citep{2024ApJ...963..139M}. \\

%From visual inspection, we found that the width of the apparent reconnection region was very small compared with the width of the magnetic islands (middle panel in Figure~6). Therefore, the approximate width of the reconnection region is $\approx$1.0 Mm. The spatial resolution of the AIA (1.5'') limits our estimation of the exact width of the current sheet.
%%%%%%%%%%%%%%%%%%%%%%%%%%%%%%%%%%%%%%%%%%%%%%%%%%%%%%%%%%%%%%%%%%%%%%%%%%%%%%%%%%%%%%%%%%%%%%%%%
\begin{figure*}
\begin{center}
  \includegraphics[trim = 0.0cm 0.0cm 0.0cm 0.0cm, scale=1.0]{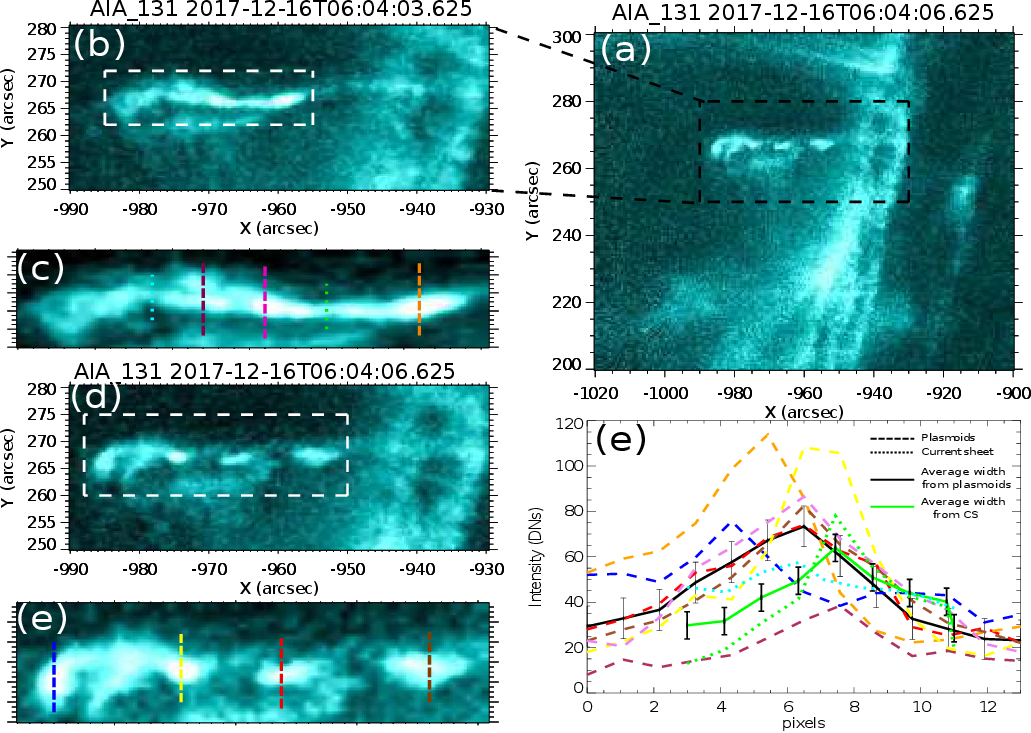}
\caption{We utilize an AIA 131 {\AA} image to illustrate the field of view (FOV) of the forced magnetic region, fragmented magnetic islands, and the associated elongated reconnection region (top-left panel) at 06:04:06 UT. The black box indicates the region of interest (ROI), which is displayed in the right vertical column at two distinct times: 06:03:06 UT and 06:04:06 UT. Two extracted sub-regions (marked by white boxes) are zoomed in and enhanced to highlight the reconnection region containing multiple magnetic islands. The bottom-right panel presents the intensity profiles across and along the elongated reconnection region for various dotted colored slits: yellow, orange, red, pink, magenta, blue, and brown slits are used to measure the width of the magnetic islands, while green and cyan slits are employed to estimate the width of the potential current sheet. The solid black curve represents the averaged fitted profile used to determine the width of the plasmoids, while the solid green curve indicates the width of the current sheet. The standard deviation is calculated and used as the error in extracting the widths of the current sheet and magnetic islands.}
\end{center}
\end{figure*}
%%%%%%%%%%%%%%%%%%%%%%%%%%%%%%%%%%%%%%%%%%%%%%%%%%%%%%%%%%%%%%%%%%%%%%%%%%%%%%%%%%%%%%%%%%%%%%%
Figure~10 shows multiple snapshots of the reconnection region in sequence between 06:00:33 UT (initiation of reconnection and formation of magnetic islands) and 06:06:21 UT (just before breaking off the apparent current sheet). The spatio-temporal evolution of the reconnection region is as follows. First, reconnection occurs inside a segment of the prominence initiated at $\sim$06:00 UT. This led to the formation of a thin elongated plasma sheet (apparent current sheet). Multiple magnetic islands or plasmoids (observables are the moving plasma blobs) were ejected bidirectionally from this elongated plasma sheet. The white arrows in panel $'a'$ of Figure~10 indicate multiple magnetic islands and their bidirectional propagation during the reconnection. During this reconnection, a chain of magnetic islands forms and is subjected to coalescence instability, as suggested by \citet{2019A&A...623A..15P}. The formation of magnetic islands and their ejection were very rapid, likewise the one recently simulated by \citet{2024ApJ...977..235M}. Here, we are not tracking each of them individually as it is a difficult process. However, we observe that once the elongated apparent current sheet forms between the two merging magnetic islands, the larger plasmoid ejects with slower velocity as compared to the smaller magnetic islands, which leads to a more rapid reconnection. \citet{2019A&A...623A..15P} suggest that the enhancement in the reconnection rate arises owing to coalescence instability. Coalescence instability is an ideal instability that evolves because of an increase in the imbalance of the Lorentz force \citep{2024ApJ...977..235M}. The merged magnetic islands propagate together as a similar current attracts \citep{2019A&A...623A..15P,2024ApJ...977..235M}. A similar scenario was observed in this study. We found that three magnetic islands formed at $\sim$06:00:33 UT (white arrow in the panel $'a'$ of Figure~10). Two nearby plasmoids merge and are ejected from the reconnection region between 06:00:33 and 06:01:33 UT. After $\sim$06:01:45 UT, another island formed, which further broke into two small islands and was ejected in the opposite direction from the reconnection region (panel $'a'$; Figure~10). These two ejected islands further merged into the nearby previously ejected plasmoid in the top and bottom parts of the plasma sheet/apparent current sheet (middle rows in panel $'a'$; Figure~10). Similar physical scenarios appeared at different spatio-temporal scales during the latter phase of the study. The ejected magnetic islands from the reconnection region merged within the top and bottom parts of the plasma sheet, confirming the onset of coalescence instability. However, we wish to clarify that this qualitative description depends on the available spatial resolution of AIA 171 {\AA}. More high-resolution and high-cadence data are required to resolve the fine-scale dynamics of individual islands and their merging processes, as recently seen in the numerical simulations \citep{2024ApJ...977..235M}. The projected length of the current sheet was $\approx$28 m before breaking the current sheet (bottom row in panel $'a'$ of Figure~10).\\

In this study, the estimated width of the apparent current sheet is $\approx$1.3 Mm (panel $'e$', Figure~9), and the length of the reconnection region/apparent current sheet is $\approx$28 Mm prior to disjointment (see panel $'a'$ of Figure~10). The estimated reconnection rate (i.e., the ratio of width and length of the current sheet) was found to be 0.045, which is very similar to the observed cases of apparent coronal current sheets \citep{2024ApJ...974..104D}. The estimated Lundquist number is $495$ if we use this reconnection rate (0.045) to maintain the Sweet-Parker reconnection scaling law. However, it should be noted that this is the lowest limit of this number just to keep the reconnection going on, and plasmoid phase may attain a much larger Lundquist number. The denser prominence system may also undergo intense radiative cooling and some thermal instability can launch the fragmentation and formation of plasmoid-like blobs even at much smaller Lundquist number as evident here \citep{2022A&A...666A..28S}. \citet{2021A&A...651A..60H} observed a bidirectional jet and the formation of an apparent current sheet/plasma sheet inside the prominence, and estimated that the Lundquist number is $\approx$200. They suggested that the lower value of the Lundquist number is either related to the projection effect or a component of magnetic reconnection that initiates the reconnection and formation of an elongated plasma sheet inside a prominence. In the present study, reconnection occurred within the cold, dense prominence plasma, with a typical plasma density of $\sim$1.0$\times$10$^{11}$ cm$^{-3}$ and a temperature range of 8000--50000 K. Oblique magnetic reconnection occurs between the eruptive loop-like eruption and contains multiple jet-like structures associated with field lines that hold the prominence segment. Therefore, both causes, such as the projection effect and components of the magnetic field reconnection, may play a significant role in reducing the lower value of the Lundquist number. After the reconnection, the plasma material was heated to 6 MK. We would like to highlight that several observational reports claim the formation of the current sheet in different magnetized structures of the solar atmosphere \citep{1998ApJ...499..934O, 2012ApJ...745L...6T, 2017ApJ...835..139S, 2019ApJ...887..137S, 2020SoPh..295..167M, 2021A&A...651A..60H, 2023ApJ...953...84M, 2024ApJ...963..139M}. In the forthcoming Subsection 3.5, we discuss the existence of plasma blob-like structures formed due to the possible co-existence of reconnection and K-H instability.\\

\subsection{Existence of Plasma Blob-Like Structures}
 Note that the reconnection region (apparent current sheet) contained multiple plasmoids (plasma blobs) before the disjoint. The other possibility of the formation of plasma blobs inside the elongated plasma sheet and the enhancement in the plasma densities are related to the Kelvin-Helmholtz (KH)/Rayleigh-Taylor (RT) instability. Several observational studies have suggested that plasma blob formation and density enhancement are related to the onset of Rayleigh-Taylor (RT) and/or Kelvin–Helmholtz (KT) instability in the prominence structure \citep{2010ApJ...716.1288B, 2011Natur.472..197B, 2010SoPh..267...75R, 2018RvMPP...2....1H, 2018ApJ...864L..10H, 2019ApJ...874...57M, 2021ApJ...923...72M} as well as in the solar corona and eruption of the jets \citep{2011ApJ...734L..11O, 2015ApJ...813..123Z, 2016SoPh..291.3165M, 2018ApJ...857..115Y, 2018NatSR...8.8136L, 2019SoPh..294...68S, 2021ApJ...923...72M}. The zoomed-in view of the plasma sheet (reconnection region) has been displayed in panel $'c'$ of Figure~10. Before disjoining the plasma sheet, we measured the separation between two consecutive plasma blobs (panel $'c'$; Figure~10). We used the cursor command available in {\it 'Solarsoft'} to extract the separation between two consecutive blobs. We measured the separation between the tip-to-tip, mid-to-mid, and bottom-to-bottom parts of the blobs. The average of these separations provides the actual separation between two consecutive plasma blobs before the disjointment of the reconnection region. The average separation was between 2.1--8 Mm (panel $'c'$; Figure~10). We also placed four slits (red, cyan, black, and green, panel $'a'$; Figure~10) to extract the width of the plasma blobs. The variation in the intensity of the AIA 171 {\AA} waveband shows a Gaussian profile. The FWHM of these Gaussian profiles provides the width of the plasma blobs. The extracted width of the plasma blobs was 7--9 pixels (3--4 Mm) (panel $'b'$; Figure~10). \\

The classical criteria for the onset of KH instability is that the shearing velocity must exceed the Alfv\'en velocity in the other layer, or the aspect ratio (width and characteristic wavelength of KH instability) must be greater than 3.5 \citep{1978SoPh...58...57P, 1981STIA...8217950D, 2018ApJ...864L..10H}. To investigate the aspect ratio, we have included two panels $'b'$ and $'c'$ in Figure~10. The wavelength of the KH unstable mode, which is the separation between two consecutive plasma blobs, was found to be between 2.1--8.0 Mm before the disconnection of the reconnection region (panel $'c'$; Figure~10). We also estimated the width of these plasma blobs using the full width at half maximum (FWHM) of the AIA 171 {\AA} intensity across the red, cyan, green, and black slits (panel 'b'; Figure~10). The estimated width was found to be between approximately 9 and 11 pixels equivalent to 3.9--4.8 Mm. The estimated aspect ratio lies between 0.5--2.1, which does not meet the classical criterion for the onset of KH instability because the aspect ratio must be greater than 3.5. Our findings, as depicted above, suggest that coalescence instability may occur as plasma blobs move bidirectionally and tend to merge. This contradicts the fastest-growing mode criteria ($\lambda$= (2-4)$\times$ $\pi$$\times$$a$) for the onset of the KH instability \citep{2018NatSR...8.8136L, 2018ApJ...864L..10H, 2021ApJ...923...72M}. Furthermore, KH instability plays a significant role in heating the localized plasma through the formation of KH unstable vortices and turbulence \citep{2019ApJ...884L..51Y}. In this study, we observed significant heating following the formation of plasma sheets and blobs. The heating occurred approximately 10 min after the formation of an elongated plasma sheet and plasma blobs (Figures~9,10 and animation2.mp4, animation3.mp4). After the formation of plasma blobs, the average temperature extracted from the box region inside the elongated plasma sheet increased upto 6 MK or even higher as seen in the derived temperature map (animation3.mp4). It is important to note that this sudden heating cannot be attributed to turbulence resulting from the KH instability. The plasma sheet was disjointed at 06:08 UT, after which the temperature began to increase and peaked at 06:12 UT (panel $'c'$ of Figure~8 and animation2.mp4, animation3.mp4). However, it should be noted that we can resolve only larger plasma blobs (bigger than 1.1 Mm similar to AIA spatial resolution, using AIA imaging observations. We might have missed some small-scale blobs that propagated inside the elongated plasma sheet/reconnection region. Therefore, the actual estimation of the separation between two plasma blobs and width estimations is difficult using imaging observations. It should be noted that we observed the relative motion of the prominence segment and loop-like eruption. In this study, we did not detect any shearing velocity or vortex formation leading to an overturning vortex. However, the relative motion between the reconnection region and surrounding prominence plasma may initiate KH instability inside elongated plasma sheets to form plasma blobs. Similar scenarios have been discussed in previous observations and simulations \citep{2013PhPl...20c2117W, 2021A&A...651A..60H, 2021ApJ...923...72M, 2023ApJ...954L..36W}. Therefore, we cannot discard the possibility of the bi-commodity of KH unstable blobs and reconnection-generated plamoids within the elongated reconnection region. Hence, it is not likely that the KH instability was the sole cause of blob formation, as the criteria for this instability were not met. Specifically, the absence of undulations or overturning vortices, the fastest-growing mode, and sudden heating persisted even after the disappearance of the plasma sheets and plasma blob. However, we cannot discard the possibility of the evolution of KH instability inside the elongated reconnection region in the presence of the relative flow of ejected blobs and surrounding cool prominence plasma. Therefore, we conclude that the most plausible mechanism for the formation of plasma blobs/magnetic islands and the subsequent heating inside the prominence segment may be the externally forced reconnection scenario, however, we can not rule out bi-commodity of reconnection and KH instability.\\%, which is triggered externally by jet-associated fields.}
%%%%%%%%%%%%%%%%%%%%%%%%%%%%%%%%%%%%%%%%%%%%%%%%%%%%%%%%%%%%%%%%%%%%%%%%%%%%%%%%%%%%%
\begin{figure*}
  \includegraphics[trim = 0.0cm 0.0cm 0.0cm 0.0cm, scale=.12]{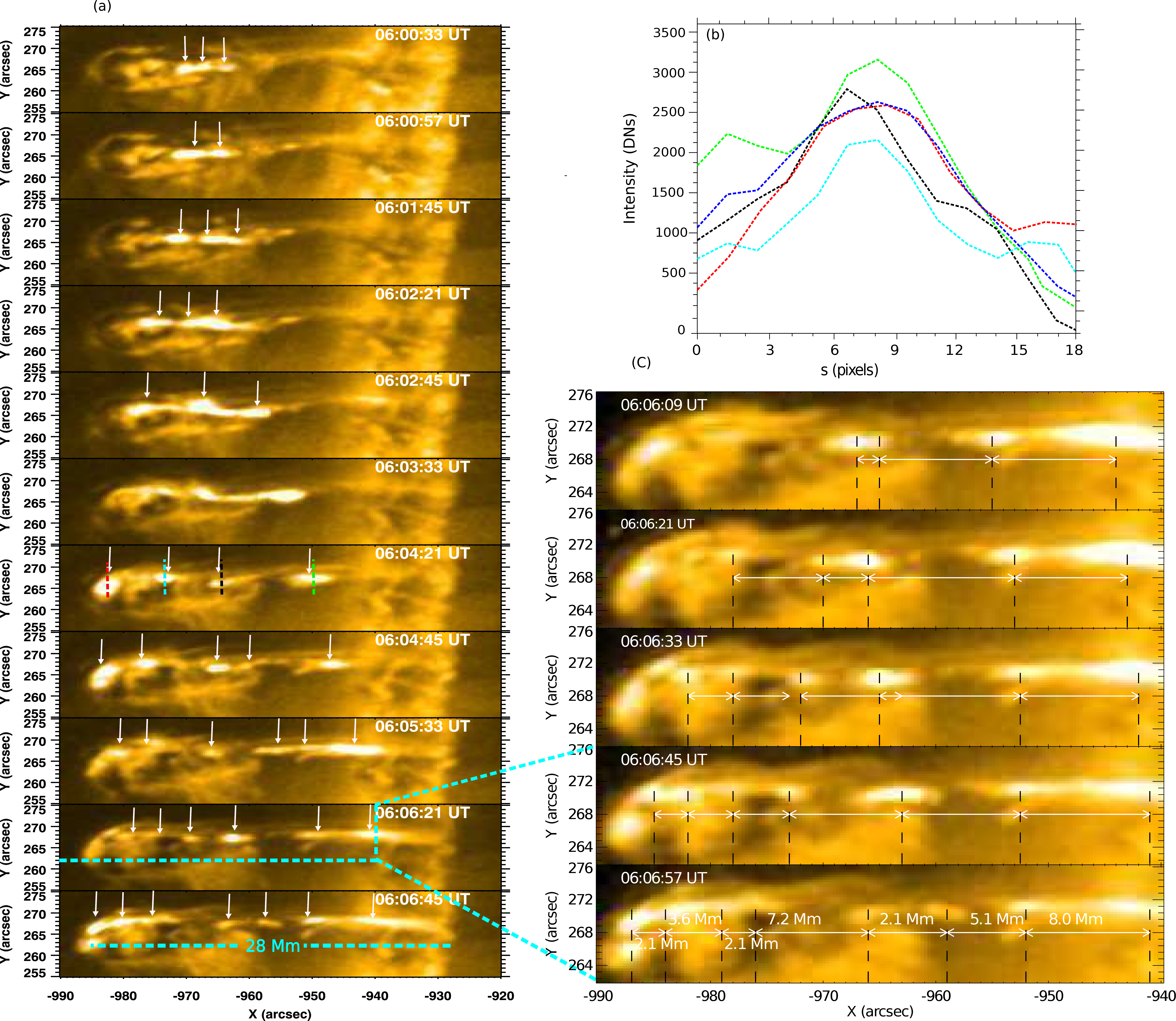}
\caption{Panel (a): zoomed in view to show the spatio-temporal evolution of the reconnection, propagation of multiple magnetic islands (indicated by white arrows), and their merging as observed in AIA 171 {\AA}. Panel (b) shows the extracted Gaussian width of the moving plasma blobs inside the elongated reconnection region from AIA 171 {\AA} waveband. Vertical black dotted lines are used to extract the separation between two consecutive moving plasma blobs inside the reconnection region (Panel (c)).}
\end{figure*}
%%%%%%%%%%%%%%%%%%%%%%%%%%%%%%%%%%%%%%%%%%%%%%%%%%%%%%%%%%%%%%%%%%%%%%%%%%%
\section{Discussion and Conclusion}

In this paper, we present the observation of a prominence segment undergoing reconnection under the influence of collimated jet-like structures continuously impinging on it. This continuous thrust of plasma coming from external sources disturbs the 
internal magnetic field of prominence and possibly misaligned various its threads, and after a time lag of 5 min, the reconnection begins. The initiation of inflow and triggering of reconnection occur when jet-like collimated structures disturb the boundary of the magnetic fields of the prominence and the field lines of the overlying flux rope. During magnetic reconnection, an elongated current sheet evolves and multiple magnetic islands here referred to as moving plasma blobs in the observational base-line are ejected along with the elongated reconnection region. Our novel analyses of observations explain the magnetic reconnection scenario caused by multiple jet-like eruptions acting as external perturbations. Our further investigation of the reconnection region reveals the formation of multiple magnetic islands inside the elongated reconnection region with a characteristic size of a few $Mm$-scale. Notably, magnetic island/plasmoid formation frequently occurs during fractional and plasmoid-induced reconnection, where the current sheet splits into several propagating plasmoids \citep{1995ApJ...451L..83S, 2001EP&S...53..473S, 2016ASSL..427..373S, 2022NatCo..13..640Y, 2024ApJ...963..139M}. Similar types of multiple plasmoid ejections have been reported for different magnetic structures of the solar atmosphere \citep{2012ApJ...745L...6T, 2018ApJ...852...98W, 2023ApJ...943..156K}. Moreover, the initiation of eruption through magnetic reconnection was investigated by employing three-dimensional simulations. This examination was conducted by considering the instability conditions and the configuration of the magnetic field. These findings indicate that the commencement of an eruption via magnetic reconnection is strongly influenced by the local structure of the magnetic field within the solar atmosphere \citep{2004ApJ...610..537K, 2012ApJ...760...31K,2024ApJ...963..139M}. In addition, the formation of magnetic islands owing to forced magnetic reconnection triggered by multiple external pulses has been established in theory and numerical modeling for solar and astrophysical plasmas \citep{1985PhFl...28.2412H, 2001PhPl....8..132B, 2005GeoRL..32.6105B, 2017JPlPh..83e2001V, 2019A&A...623A..15P, 2005PhPl...12a2904J}. However, observationally, the formation of magnetic islands inside an elongated current sheet within a forced reconnection region has not been reported before \citep{2010ApJ...712L.111J, 2019ApJ...887..137S, 2020A&A...643A.140M, 2021ApJ...920...18S}. Our result shows that multiple magnetic islands propagate and merge within each other in the top and lower parts of the prominence, as suggested by \citet{2019A&A...623A..15P} and \citet{2024ApJ...977..235M}. They also suggest that the distinguishing trait of forced reconnection is the formation of magnetic islands inside the elongated reconnection region that subsequently merges, increasing the reconnection rate. A similar physical scenario was observed in the present study. The estimated reconnection rate increases periodically as the forced reconnection proceeds. The merger of the islands leads to 
liberation of thermal energy (5.4$\times$$\approx$10$^{27}$ erg) in the solar corona and is responsible for the heating of the 
cool and dense plasma. \citet{2021ApJ...908..213L} estimated the thermal energy associated with current sheet and filament eruptions. The estimated thermal energies from these two regions were $\approx$10$^{25}$ erg (current sheet) and $\approx$10$^{34}$ erg (during filament eruption), respectively. The thermal energy estimated using the present observation is of the same order as that diagnosed by \citep{2021ApJ...908..213L}, which is significant for heating the prominence plasma to the coronal temperature. We have diagnosed the strength of reconnecting field lines to be 29--56 G. DEM analysis and the temporal evolution of EM also suggest that significant heating occurs after the merging of magnetic islands, releasing 
stored magnetic energy. \\

%We also estimate the magnetic diffusitivity and current density from the reconnection region. The value of the estimated magnetic diffusitivity has highly depend upon the way of the calculation. The estimated magnetic diffusivity in the reconnection region is $\approx$10$^{2}$--10$^{4}$ m$^{2}$ s$^{-1}$ (using reconnection rate), while it is found to be 0.25-0.50 m$^{2}$ s$^{-1}$ (using coronal Lundquist number) is much smaller for the onset of spontaneous reconnection in the cool and dense prominence plasma. A significant current density (2--5$\times$10$^{-4}$ A m$^{-2}$) also appears that causes the drainage of cool prominence plasma and heat it up to coronal temperature.\\
As noted in the present observation, external disturbances may perturb the boundary of the magnetized plasma to initiate the formation of the current sheet and the reconnection region. Release of a large amount of magnetic energy 
following small-energy external perturbations is a crucial feature of forced magnetic reconnection \citep{2005PhPl...12a2904J}. In the present observations, the collimated jets may act as an external perturbation to disturb the solar prominence-corona boundary and initiate forced reconnection. This paper also explains the kinematics of the reconnection region, its energetics, and its thermal structure. A substantial amount of thermal energy is liberated during reconnection. The highly dynamic nature of the solar atmosphere supports the initiation of the forced magnetic reconnection, which may be a feasible candidate for solar coronal heating \citep{2005PhPl...12a2904J, 2017JPlPh..83e2001V, 2019A&A...623A..15P}. The first observational evidence of forced magnetic reconnection in the solar corona \citep{2019ApJ...887..137S}, it has been suggested that the dynamic nature of the solar atmosphere may possess such a reconnection scenario on different spatiotemporal scales. There are few observations of forced reconnections recently reported in the solar atmosphere \citep{2010ApJ...712L.111J, 2019ApJ...887..137S, 2020A&A...643A.140M, 2021ApJ...920...18S}. The observations of reconnection presented and analyzed here demonstrate the fine dynamics (magnetic island properties/plasmoids), energetics, and heating scenarios caused by external perturbations on a prominence magnetic field system due to multiple coronal jet-like structures (similar to the case of forced reconnection), which are ubiquitous.\\

The release of a large amount of magnetic energy after accumulating small amounts of energy as external perturbations is a crucial feature of forced magnetic reconnection \citep{2005PhPl...12a2904J}. A major question that needs to be answered is: Why is forced reconnection not observed everywhere in a dynamic solar atmosphere? One major limitation of observing forced reconnection is the line of sight and orientation of the magnetic field. Another issue is that forced reconnection occurs on a small scale, with a rapid reconnection rate. Therefore, more detailed observations are needed to explore forced reconnection using existing and future high-resolution ground- and space-based observations. Further observations are required for a more detailed analysis of the forced reconnection region, its heating mechanism, and its magnetic structure in the solar atmosphere at different spatio-temporal scales using high-resolution imaging and spectroscopic data of existing space instruments (AIA, IRIS, Solar Orbiter, DKIST, etc.) as well as upcoming space instruments such as SUIT and VELC/ADITYA-L1. In addition, high-resolution observations using new instruments at the SST and DKIST are useful for understanding the onset of forced magnetic reconnection from the lower solar atmosphere to the lower corona. Understanding such exclusive physical processes of the 'forced reconnection' is highly required both in observations and modeling, as one step ahead of this concept the Symbiosis of WAves \& Reconnection is conceptualized recently \citep{Sri24}, and its one example is seen in the numerical modeling \citep{2024ApJ...977..235M} in solar-like plasma, and possibly in recent experiments at the laboratory scales also \citep{2024Symm...16..103F}. \\

We also discuss alternative possibilities for the formation of plasma blobs and enhancement in plasma densities, which are governed by the Kelvin-Helmholtz/Rayeligh-Taylor instability \citep{2013PhPl...20c2117W, 2017ApJ...841...27N, 2018NatSR...8.8136L, 2021A&A...651A..60H, 2021ApJ...923...72M, 2022ApJ...931L..32W, 2023ApJ...954L..36W}. However, we found that the formed plasma blobs did not satisfy the fastest-growing mode criteria for the onset of the KH instability, and a sudden enhancement in the heating after the appearance of the plasma sheet confirms that the only KH instability is not a likely physical mechanism for the present observation. By applying the forced reconnection model in the presence of multiple pulses and the formation of multiple magnetic islands, bidirectional flows, and sudden heating after the appearance of the plasma sheet, the enhancement in density suggests that magnetic reconnection is one of the plausible mechanism to explain the formation of the plasma blobs inside the elongated plasma sheet/current sheet. Also during the formation of the elongated reconnection region and propagation of multiple plasma blobs, a relative velocity or shear flow may develop within the reconnection region, which can initiate the KH instability and may contribute to the formation of bigger blobs/plasmoids \citep{2021A&A...651A..60H, 2023ApJ...954L..36W}. Therefore, we conclude that both reconnection and KH instability and their coupled behavior are important to understand such dynamical phenomena inside the elongated reconnection region. The future detailed study can understand this possible co-existence of reconnection region and KH unstable plasma blobs using a large cohort of high-resolution observations, and it should be considered yet as an open problem.\\

\section{Acknowledgements}
S. K. Mishra acknowledges the Kyoto University, Japan, and Indian Institute of Astrophysics (IIA, Bangalore) for providing him institute fellowship and computational facilities. S. K. Mishra is grateful to Professor Ayumi Asai, Kyoto University, for the scientific discussion and her fruitful suggestions. A.K.S. acknowledges the ISRO Project Grant (DS\_2B-13012(2)/26/2022-Sec.2) for the support of his research. S.P.R. acknowledges support from SERB (Govt. of India) research grant CRG/2019/003786. P.J. work was sponsored by the DynaSun project and has thus received funding under the Horizon Europe programme of the European Union under grant agreement (no. 101131534). Views and opinions expressed are however those of the author(s) only and do not necessarily reflect those of the European Union and therefore the European Union cannot be held responsible for them. We acknowledge the use of data from Cheung et al. (2015) to calculate the differential emission measure (DEM). Data are courtesy of NASA/SDO and the AIA science team.


\begin{thebibliography}{}
\bibitem[Antolin et al.(2021)]{2021NatAs...5...54A} Antolin, P., Pagano, P., Testa, P., et al.\ 2021, Nature Astronomy, 5, 54. doi:10.1038/s41550-020-1199-8
 \bibitem[Aschwanden(2004)]{2004psci.book.....A} Aschwanden, M.~J.\ 2004, Physics of the Solar Corona. An Introduction, by M.J. Aschwanden. Published by Praxis Publishing Ltd., Chichester, UK, and Springer-Verlag Berlin ISBN..3-540-22321-5, 2004.
 \bibitem[Berghmans et al.(2021)]{2021A&A...656L...4B} Berghmans, D., Auch{\`e}re, F., Long, D.~M., et al.\ 2021, \aap, 656, L4. doi:10.1051/0004-6361/202140380
\bibitem[Berger et al.(2010)]{2010ApJ...716.1288B} Berger, T.~E., Slater, G., Hurlburt, N., et al.\ 2010, \apj, 716, 1288 
\bibitem[Berger et al.(2011)]{2011Natur.472..197B} Berger, T., Testa, P., Hillier, A., et al.\ 2011, \nat, 472, 197 
\bibitem[Berger et al.(2017)]{2017ApJ...850...60B} Berger, T., Hillier, A., \& Liu, W.\ 2017, \apj, 850, 60 
\bibitem[Birn et al.(2005)]{2005GeoRL..32.6105B} Birn, J., Galsgaard, K., Hesse, M., et al.\ 2005, \grl, 32, L06105
\bibitem[Browning et al.(2001)]{2001PhPl....8..132B} Browning, P.~K., Kawaguchi, J., Kusano, K., et al.\ 2001, Physics of Plasmas, 8, 132
\bibitem[Chen et al.(2001)]{2001ChJAA...1..176C} Chen, P.-F., Fang, C., \& Ding, M.-D.~D.\ 2001, \cjaa, 1, 176 
\bibitem[Chen \& Ding(2006)]{2006ApJ...641.1217C} Chen, Q.~R., \& Ding, M.~D.\ 2006, \apj, 641, 1217
\bibitem[Chen et al.(2004)]{2004ApJ...602L..61C} Chen, P.~F., Shibata, K., Brooks, D.~H., et al.\ 2004, \apjl, 602, L61. doi:10.1086/382479
\bibitem[Chitta et al.(2021)]{2021A&A...647A.159C} Chitta, L.~P., Peter, H., \& Young, P.~R.\ 2021, \aap, 647, A159. doi:10.1051/0004-6361/202039969
 \bibitem[Chen(2011)]{2011LRSP....8....1C} Chen, P.~F.\ 2011, Living Reviews in Solar Physics, 8, 1. doi:10.12942/lrsp-2011-1
\bibitem[Chen et al.(2019)]{2019ApJ...879...74C} Chen, H., Yang, J., Duan, Y., et al.\ 2019, \apj, 879, 74. doi:10.3847/1538-4357/ab24ce
\bibitem[Cheung \& Isobe(2014)]{2014LRSP...11....3C} Cheung, M.~C.~M. \& Isobe, H.\ 2014, Living Reviews in Solar Physics, 11, 3. doi:10.12942/lrsp-2014-3
\bibitem[Cheung et al.(2015)]{2015ApJ...807..143C} Cheung, M.~C.~M., Boerner, P., Schrijver, C.~J., et al.\ 2015, \apj, 807, 143 
\bibitem[Cirtain et al.(2013)]{2013Natur.493..501C} Cirtain, J.~W., Golub, L., Winebarger, A.~R., et al.\ 2013, \nat, 493, 501. doi:10.1038/nature11772
\bibitem[Comisso et al.(2015)]{2015PhPl...22d2109C} Comisso, L., Grasso, D., \& Waelbroeck, F.~L.\ 2015, Physics of Plasmas, 22, 042109. doi:10.1063/1.4918331
\bibitem[De Pontieu et al.(2014)]{2014SoPh..289.2733D} De Pontieu, B., Title, A.~M., Lemen, J.~R., et al.\ 2014, \solphys, 289, 2733. doi:10.1007/s11207-014-0485-y
\bibitem[Drazin \& Reid(1981)]{1981STIA...8217950D} Drazin, P.~G. \& Reid, W.~H.\ 1981, NASA STI/Recon Technical Report A, 82, 17950
\bibitem[Ding \& Zhang(2024)]{2024ApJ...974..104D} Ding, T. \& Zhang, J.\ 2024, \apj, 974, 104. doi:10.3847/1538-4357/ad6df5
\bibitem[Ding et al.(2011)]{2011A&A...535A..95D} Ding, J.~Y., Madjarska, M.~S., Doyle, J.~G., et al.\ 2011, \aap, 535, A95. doi:10.1051/0004-6361/201117515
\bibitem[Hara et al.(2006)]{2006ApJ...648..712H} Hara, H., Nishino, Y., Ichimoto, K., et al.\ 2006, \apj, 648, 712. doi:10.1086/505638
\bibitem[Foullon et al.(2011)]{2011ApJ...729L...8F} Foullon, C., Verwichte, E., Nakariakov, V.~M., et al.\ 2011, \apjl, 729, L8. doi:10.1088/2041-8205/729/1/L8
\bibitem[Foullon et al.(2013)]{2013ApJ...767..170F} Foullon, C., Verwichte, E., Nykyri, K., et al.\ 2013, \apj, 767, 170. doi:10.1088/0004-637X/767/2/170
\bibitem[Frank \& Savinov(2024)]{2024Symm...16..103F} Frank, A.~G. \& Savinov, S.~A.\ 2024, Symmetry, 16, 103. doi:10.3390/sym16010103
 
\bibitem[Hesse \& Cassak(2020)]{2020JGRA..12525935H} Hesse, M. \& Cassak, P.~A.\ 2020, Journal of Geophysical Research (Space Physics), 125, e25935. doi:10.1029/2018JA025935
\bibitem[Hillier \& Polito(2021)]{2021A&A...651A..60H} Hillier, A. \& Polito, V.\ 2021, \aap, 651, A60. doi:10.1051/0004-6361/201935774
\bibitem[Hillier(2018)]{2018RvMPP...2....1H} Hillier, A.\ 2018, Reviews of Modern Plasma Physics, 2, 1. doi:10.1007/s41614-017-0013-2
\bibitem[Hillier \& Polito(2018)]{2018ApJ...864L..10H} Hillier, A. \& Polito, V.\ 2018, \apjl, 864, L10. doi:10.3847/2041-8213/aad9a5
\bibitem[Hahm \& Kulsrud(1985)]{1985PhFl...28.2412H} Hahm, T.~S., \& Kulsrud, R.~M.\ 1985, Physics of Fluids, 28, 2412
\bibitem[Inhester(2006)]{2006astro.ph.12649I} Inhester, B.\ 2006, astro-ph/0612649. doi:10.48550/arXiv.astro-ph/0612649

\bibitem[Innes et al.(1997)]{1997Natur.386..811I} Innes, D.~E., Inhester, B., Axford, W.~I., et al.\ 1997, \nat, 386, 811
\bibitem[Jain et al.(2005)]{2005PhPl...12a2904J} Jain, R., Browning, P., \& Kusano, K.\ 2005, Physics of Plasmas, 12, 012904
\bibitem[Jess et al.(2010)]{2010ApJ...712L.111J} Jess, D.~B., Mathioudakis, M., Browning, P.~K., et al.\ 2010, \apjl, 712, L111
\bibitem[Jel{\'\i}nek et al.(2015)]{2015A&A...581A.131J} Jel{\'\i}nek, P., Srivastava, A.~K., Murawski, K., et al.\ 2015, \aap, 581, A131. doi:10.1051/0004-6361/201424234
\bibitem[Joshi et al.(2020)]{2020A&A...641L...5J} Joshi, J., Rouppe van der Voort, L.~H.~M., \& de la Cruz Rodr{\'\i}guez, J.\ 2020, \aap, 641, L5. doi:10.1051/0004-6361/202038769
 \bibitem[Khabarova et al.(2021)]{2021SSRv..217...38K} Khabarova, O., Malandraki, O., Malova, H., et al.\ 2021, \ssr, 217, 38. doi:10.1007/s11214-021-00814-x
\bibitem[Kusano et al.(2012)]{2012ApJ...760...31K} Kusano, K., Bamba, Y., Yamamoto, T.~T., et al.\ 2012, \apj, 760, 31. doi:10.1088/0004-637X/760/1/31
\bibitem[Kusano et al.(2004)]{2004ApJ...610..537K} Kusano, K., Maeshiro, T., Yokoyama, T., et al.\ 2004, \apj, 610, 537. doi:10.1086/421547
\bibitem[Labrosse et al.(2010)]{2010SSRv..151..243L} Labrosse, N., Heinzel, P., Vial, J.-C., et al.\ 2010, \ssr, 151, 243. doi:10.1007/s11214-010-9630-6
\bibitem[Lemen et al.(2012)]{2012SoPh..275...17L} Lemen, J.~R., Title, A.~M., Akin, D.~J., et al.\ 2012, \solphys, 275, 17
\bibitem[Lin et al.(2005)]{2005ApJ...622.1251L} Lin, J., Ko, Y.-K., Sui, L., et al.\ 2005, \apj, 622, 1251. doi:10.1086/428110
\bibitem[Li et al.(2016)]{2016NatPh..12..847L} Li, L., Zhang, J., Peter, H., et al.\ 2016, Nature Physics, 12, 847
\bibitem[Li et al.(2018)]{2018NatSR...8.8136L} Li, X., Zhang, J., Yang, S., et al.\ 2018, Scientific Reports, 8, 8136. doi:10.1038/s41598-018-26581-4
\bibitem[Li et al.(2021)]{2021ApJ...908..213L} Li, L., Peter, H., Chitta, L.~P., et al.\ 2021, \apj, 908, 213. doi:10.3847/1538-4357/abd47e
\bibitem[Litvinenko(1999)]{1999ApJ...515..435L} Litvinenko, Y.~E.\ 1999, \apj, 515, 435 
\bibitem[Mackay et al.(2010)]{2010SSRv..151..333M} Mackay, D.~H., Karpen, J.~T., Ballester, J.~L., et al.\ 2010, \ssr, 151, 333. doi:10.1007/s11214-010-9628-0
\bibitem[Masuda et al.(1994)]{1994Natur.371..495M} Masuda, S., Kosugi, T., Hara, H., et al.\ 1994, \nat, 371, 495
\bibitem[Mishra \& Srivastava(2019)]{2019ApJ...874...57M} Mishra, S.~K., \& Srivastava, A.~K.\ 2019, \apj, 874, 57
\bibitem[Mishra et al.(2018)]{2018ApJ...856...86M} Mishra, S.~K., Singh, T., Kayshap, P., et al.\ 2018, \apj, 856, 86. doi:10.3847/1538-4357/aaae03
\bibitem[Mishra et al.(2020)]{2020SoPh..295..167M} Mishra, S.~K., Srivastava, A.~K., \& Chen, P.~F.\ 2020, \solphys, 295, 167. doi:10.1007/s11207-020-01733-w
 \bibitem[Mishra et al.(2021)]{2021ApJ...923...72M} Mishra, S.~K., Singh, B., Srivastava, A.~K., et al.\ 2021, \apj, 923, 72. doi:10.3847/1538-4357/ac2a43
\bibitem[Mishra et al.(2023)]{2023ApJ...945..113M} Mishra, S.~K., Sangal, K., Kayshap, P., et al.\ 2023, \apj, 945, 113. doi:10.3847/1538-4357/acb058
\bibitem[Mishin \& Tomozov(2016)]{2016SoPh..291.3165M} Mishin, V.~V. \& Tomozov, V.~M.\ 2016, \solphys, 291, 3165. doi:10.1007/s11207-016-0891-4
\bibitem[M{\'e}sz{\'a}rosov{\'a} \& G{\"o}m{\"o}ry(2020)]{2020A&A...643A.140M} M{\'e}sz{\'a}rosov{\'a}, H. \& G{\"o}m{\"o}ry, P.\ 2020, \aap, 643, A140. doi:10.1051/0004-6361/202038388
\bibitem[M{\"o}stl et al.(2013)]{2013ApJ...766L..12M} M{\"o}stl, U.~V., Temmer, M., \& Veronig, A.~M.\ 2013, \apjl, 766, L12. doi:10.1088/2041-8205/766/1/L12
\bibitem[Mondal et al.(2023)]{2023ApJ...953...84M} Mondal, S., Srivastava, A.~K., Mishra, S.~K., et al.\ 2023, \apj, 953, 84. doi:10.3847/1538-4357/acd2da
\bibitem[Mondal et al.(2024a)]{2024ApJ...963..139M} Mondal, S., Srivastava, A.~K., Pontin, D.~I., et al.\ 2024, \apj, 963, 139. doi:10.3847/1538-4357/ad2079
%\bibitem[Mondal et al.(2024b)]{2024ApJ...977..235M} Mondal, S., Srivastava, A.~K., Pontin, D.~I., et al.\ 2024, arXiv:2411.02180. doi:10.48550/arXiv.2411.02180
\bibitem[Mondal et al.(2024b)]{2024ApJ...977..235M} Mondal, S., Srivastava, A.~K., Pontin, D.~I., et al.\ 2024, \apj, 977, 235. doi:10.3847/1538-4357/ad9022
\bibitem[Mondal et al.(2025)]{2025ApJ...979..207M} Mondal, S., Bairagi, A., \& Srivastava, A.~K.\ 2025, \apj, 979, 207. doi:10.3847/1538-4357/ada1d6
\bibitem[M{\"u}ller et al.(2020)]{2020A&A...642A...1M} M{\"u}ller, D., St. Cyr, O.~C., Zouganelis, I., et al.\ 2020, \aap, 642, A1. doi:10.1051/0004-6361/202038467
\bibitem[Ohyama \& Shibata(1998)]{1998ApJ...499..934O} Ohyama, M. \& Shibata, K.\ 1998, \apj, 499, 934. doi:10.1086/305652
\bibitem[Ofman \& Thompson(2011)]{2011ApJ...734L..11O} Ofman, L. \& Thompson, B.~J.\ 2011, \apjl, 734, L11. doi:10.1088/2041-8205/734/1/L11
\bibitem[Ni et al.(2017)]{2017ApJ...841...27N} Ni, L., Zhang, Q.-M., Murphy, N.~A., et al.\ 2017, \apj, 841, 27. doi:10.3847/1538-4357/aa6ffe
 \bibitem[Kumar et al.(2023)]{2023ApJ...943..156K} Kumar, P., Karpen, J.~T., Antiochos, S.~K., et al.\ 2023, \apj, 943, 156. doi:10.3847/1538-4357/acaea4
\bibitem[Parker(1988)]{1988ApJ...330..474P} Parker, E.~N.\ 1988, \apj, 330, 474. doi:10.1086/166485
\bibitem[Parenti(2014)]{2014LRSP...11....1P} Parenti, S.\ 2014, Living Reviews in Solar Physics, 11, 1. doi:10.12942/lrsp-2014-1
\bibitem[Peter et al.(2014)]{2014Sci...346C.315P} Peter, H., Tian, H., Curdt, W., et al.\ 2014, Science, 346, 1255726
\bibitem[Pesnell et al.(2012)]{2012SoPh..275....3P} Pesnell, W.~D., Thompson, B.~J., \& Chamberlin, P.~C.\ 2012, \solphys, 275, 3. doi:10.1007/s11207-011-9841-3
\bibitem[Potter et al.(2019)]{2019A&A...623A..15P} Potter, M.~A., Browning, P.~K., \& Gordovskyy, M.\ 2019, \aap, 623, A15
%\bibitem[Priest \& Forbes(1986)]{1986JGR....91.5579P} Priest, E.~R. \& Forbes, T.~G.\ 1986, \jgr, 91, 5579
 \bibitem[Priest \& Forbes(2007)]{2007mare.book.....P} Priest, E., \& Forbes, T.\ 2007, Magnetic Reconnection, by Eric Priest , Terry Forbes, Cambridge, UK: Cambridge
\bibitem[Priest \& Forbes(2000)]{2000mrmt.conf.....P} Priest, E. \& Forbes, T.\ 2000, Magnetic reconnection : MHD theory and applications / Eric Priest
\bibitem[Priest(1978)]{1978SoPh...58...57P} Priest, E.~R.\ 1978, \solphys, 58, 57. doi:10.1007/BF00152555
\bibitem[Ryutova et al.(2010)]{2010SoPh..267...75R} Ryutova, M., Berger, T., Frank, Z., Tarbell, T., \& Title, A.\ 2010, \solphys, 267, 75
\bibitem[Ruan et al.(2018)]{2018A&A...618A.135R} Ruan, W., Xia, C., \& Keppens, R.\ 2018, \aap, 618, A135. doi:10.1051/0004-6361/201833362
 \bibitem[Shibata(1999)]{1999Ap&SS.264..129S} Shibata, K.\ 1999, \apss, 264, 129. doi:10.1023/A:1002413214356
\bibitem[Shibata et al.(1995)]{1995ApJ...451L..83S} Shibata, K., Masuda, S., Shimojo, M., et al.\ 1995, \apjl, 451, L83. doi:10.1086/309688
\bibitem[Rouppe van der Voort et al.(2021)]{2021A&A...648A..54R} Rouppe van der Voort, L.~H.~M., Joshi, J., Henriques, V.~M.~J., et al.\ 2021, \aap, 648, A54. doi:10.1051/0004-6361/202040171
 \bibitem[Rochus et al.(2020)]{2020A&A...642A...8R} Rochus, P., Auch{\`e}re, F., Berghmans, D., et al.\ 2020, \aap, 642, A8. doi:10.1051/0004-6361/201936663
\bibitem[Samanta et al.(2019)]{2019Sci...366..890S} Samanta, T., Tian, H., Yurchyshyn, V., et al.\ 2019, Science, 366, 890. doi:10.1126/science.aaw2796
\bibitem[Scharmer et al.(2008)]{2008ApJ...689L..69S} Scharmer, G.~B., Narayan, G., Hillberg, T., et al.\ 2008, \apjl, 689, L69. doi:10.1086/595744
\bibitem[Sen \& Keppens(2022)]{2022A&A...666A..28S} Sen, S. \& Keppens, R.\ 2022, \aap, 666, A28. doi:10.1051/0004-6361/202244152
\bibitem[Shibata et al.(1992)]{1992PASJ...44..265S} Shibata, K., Nozawa, S., \& Matsumoto, R.\ 1992, \pasj, 44, 265
\bibitem[Shibata \& Tanuma(2001)]{2001EP&S...53..473S} Shibata, K. \& Tanuma, S.\ 2001, Earth, Planets and Space, 53, 473. doi:10.1186/BF03353258
\bibitem[Shibata et al.(2007)]{2007Sci...318.1591S} Shibata, K., Nakamura, T., Matsumoto, T., et al.\ 2007, Science, 318, 1591. doi:10.1126/science.1146708
\bibitem[Shibata \& Magara(2011)]{2011LRSP....8....6S} Shibata, K. \& Magara, T.\ 2011, Living Reviews in Solar Physics, 8, 6. doi:10.12942/lrsp-2011-6
\bibitem[Shibata \& Takasao(2016)]{2016ASSL..427..373S} Shibata, K. \& Takasao, S.\ 2016, Magnetic Reconnection: Concepts and Applications, 427, 373. doi:10.1007/978-3-319-26432-5\_10
\bibitem[Solanki et al.(2019)]{2019SoPh..294...68S} Solanki, R., Srivastava, A.~K., Rao, Y.~K., et al.\ 2019, \solphys, 294, 68. doi:10.1007/s11207-019-1453-3
\bibitem[Srivastava et al.(2019)]{2019ApJ...887..137S} Srivastava, A.~K., Mishra, S.~K., Jel{\'\i}nek, P., et al.\ 2019, \apj, 887, 137
\bibitem[Srivastava et al.(2021)]{2021ApJ...920...18S} Srivastava, A.~K., Mishra, S.~K., \& Jel{\'\i}nek, P.\ 2021, \apj, 920, 18. doi:10.3847/1538-4357/ac1519
\bibitem[Srivastava et al.(2024)]{Sri24} Srivastava, A.~K., Priest, E.~R., Ofman, L., Mondal, Sripan, Kwon, R.-Y., Pontin, D., Murawski, K., Mishra, S.~K., Yuan, Ding, Asai, A., 2024, COSPAR 44th Scientific Assembly, Busan, S. Korea, E2.7-0002-24.
\bibitem[Seaton et al.(2017)]{2017ApJ...835..139S} Seaton, D.~B., Bartz, A.~E., \& Darnel, J.~M.\ 2017, \apj, 835, 139. doi:10.3847/1538-4357/835/2/139
\bibitem[Sterling et al.(2015)]{2015Natur.523..437S} Sterling, A.~C., Moore, R.~L., Falconer, D.~A., et al.\ 2015, \nat, 523, 437
\bibitem[Su et al.(2013)]{2013NatPh...9..489S} Su, Y., Veronig, A.~M., Holman, G.~D., et al.\ 2013, Nature Physics, 9, 489. doi:10.1038/nphys2675
\bibitem[Sun et al.(2015)]{2015NatCo...6.7598S} Sun, J.~Q., Cheng, X., Ding, M.~D., et al.\ 2015, Nature Communications, 6, 7598
\bibitem[Tian et al.(2014)]{2014Sci...346A.315T} Tian, H., DeLuca, E.~E., Cranmer, S.~R., et al.\ 2014, Science, 346, 1255711
\bibitem[Testa et al.(2014)]{2014Sci...346B.315T} Testa, P., De Pontieu, B., Allred, J., et al.\ 2014, Science, 346, 1255724. doi:10.1126/science.1255724
\bibitem[Takasao et al.(2012)]{2012ApJ...745L...6T} Takasao, S., Asai, A., Isobe, H., et al.\ 2012, \apjl, 745, L6. doi:10.1088/2041-8205/745/1/L6
 \bibitem[van Driel-Gesztelyi et al.(2014)]{2014ApJ...788...85V} van Driel-Gesztelyi, L., Baker, D., T{\"o}r{\"o}k, T., et al.\ 2014, \apj, 788, 85. doi:10.1088/0004-637X/788/1/85
\bibitem[Vekstein \& Jain(1998)]{1998PhPl....5.1506V} Vekstein, G.~E., \& Jain, R.\ 1998, Physics of Plasmas, 5, 1506University Press, 2007
\bibitem[Vekstein \& Kusano(2015)]{2015PhPl...22i0707V} Vekstein, G. \& Kusano, K.\ 2015, Physics of Plasmas, 22, 090707. doi:10.1063/1.4932079
\bibitem[Vekstein(2017)]{2017JPlPh..83e2001V} Vekstein, G.\ 2017, Journal of Plasma Physics, 83, 205830501
\bibitem[Xue et al.(2016)]{2016NatCo...711837X} Xue, Z., Yan, X., Cheng, X., et al.\ 2016, Nature Communications, 7, 11837
\bibitem[Xue et al.(2018)]{2018ApJ...858L...4X} Xue, Z., Yan, X., Yang, L., et al.\ 2018, \apjl, 858, L4. doi:10.3847/2041-8213/aabe77
\bibitem[Wang \& Bhattacharjee(1992)]{1992PhFlB...4.1795W} Wang, X. \& Bhattacharjee, A.\ 1992, Physics of Fluids B, 4, 1795. doi:10.1063/1.860035
\bibitem[Wang et al.(2017)]{2017NatCo...8.1330W} Wang, W., Liu, R., Wang, Y., et al.\ 2017, Nature Communications, 8, 1330
 \bibitem[Wang et al.(2022)]{2022ApJ...931L..32W} Wang, Y., Cheng, X., Ren, Z., et al.\ 2022, \apjl, 931, L32. doi:10.3847/2041-8213/ac715a
 \bibitem[Wang et al.(2023)]{2023ApJ...954L..36W} Wang, Y., Cheng, X., Ding, M., et al.\ 2023, \apjl, 954, L36. doi:10.3847/2041-8213/acf19d
\bibitem[Warren et al.(2018)]{2018ApJ...854..122W} Warren, H.~P., Brooks, D.~H., Ugarte-Urra, I., et al.\ 2018, \apj, 854, 122. doi:10.3847/1538-4357/aaa9b8
\bibitem[Wuelser et al.(2004)]{2004SPIE.5171..111W} Wuelser, J.-P., Lemen, J.~R., Tarbell, T.~D., et al.\ 2004, \procspie, 5171, 111. doi:10.1117/12.506877
\bibitem[Wyper \& Pontin(2013)]{2013PhPl...20c2117W} Wyper, P.~F. \& Pontin, D.~I.\ 2013, Physics of Plasmas, 20, 032117. doi:10.1063/1.4798516
\bibitem[Wyper et al.(2018)]{2018ApJ...852...98W} Wyper, P.~F., DeVore, C.~R., \& Antiochos, S.~K.\ 2018, \apj, 852, 98. doi:10.3847/1538-4357/aa9ffc
\bibitem[Yan et al.(2022)]{2022NatCo..13..640Y} Yan, X., Xue, Z., Jiang, C., et al.\ 2022, Nature Communications, 13, 640. doi:10.1038/s41467-022-28269-w
\bibitem[Yamada et al.(2010)]{2010RvMP...82..603Y} Yamada, M., Kulsrud, R., \& Ji, H.\ 2010, Reviews of Modern Physics, 82, 603
 \bibitem[Yang et al.(2018)]{2018ApJ...857..115Y} Yang, H., Xu, Z., Lim, E.-K., et al.\ 2018, \apj, 857, 115. doi:10.3847/1538-4357/aab789
\bibitem[Yokoyama \& Shibata(1995)]{1995Natur.375...42Y} Yokoyama, T. \& Shibata, K.\ 1995, \nat, 375, 42
\bibitem[Yokoyama \& Shibata(1996)]{1996PASJ...48..353Y} Yokoyama, T. \& Shibata, K.\ 1996, \pasj, 48, 353. doi:10.1093/pasj/48.2.353
 \bibitem[Yuan et al.(2019)]{2019ApJ...884L..51Y} Yuan, D., Shen, Y., Liu, Y., et al.\ 2019, \apjl, 884, L51. doi:10.3847/2041-8213/ab4bcd
\bibitem[Zaqarashvili et al.(2015)]{2015ApJ...813..123Z} Zaqarashvili, T.~V., Zhelyazkov, I., \& Ofman, L.\ 2015, \apj, 813, 123. doi:10.1088/0004-637X/813/2/123
 \bibitem[Zhelyazkov et al.(2015)]{2015A&A...574A..55Z} Zhelyazkov, I., Zaqarashvili, T.~V., \& Chandra, R.\ 2015, \aap, 574, A55. doi:10.1051/0004-6361/201424793
\bibitem[Zhang et al.(2012)]{2012NatCo...3..747Z} Zhang, J., Cheng, X., \& Ding, M.-D.\ 2012, Nature Communications, 3, 747


\end{thebibliography}
\end{document}